\documentclass[12pt]{article}
\usepackage{graphicx,amsmath,amssymb,cite}
\usepackage{epsfig,epsf}
\usepackage{graphicx}
\usepackage{epstopdf}
\newcommand{\beq}{\begin{equation}}
\newcommand{\eeq}{\end{equation}}
\newcommand{\alphabar}{\bar{\alpha_s}}
\newcommand{\betabar}{\bar{\beta_0}}
\newcommand{\factor}{\left[ \frac{\Gamma(1/2+i\nu)}{\Gamma(1/2-i\nu)}e^{-2i\Psi(1) \nu}\right]}

\numberwithin{equation}{section}

\begin{document}

\begin{flushright}
DESY 12-085\\
SHEP - 12-12\\
\end{flushright}

\vspace*{0.5 cm}

\begin{center}

{\Large{\bf BFKL Evolution as a Communicator Between  Small and Large Energy Scales  }}

\vspace*{1 cm}

{\large H. Kowalski~$^1$, L.N. Lipatov~$^{2}$, and D.A. Ross~$^3$} \\ [0.5cm]
{\it $^1$ Deutsches Elektronen-Synchrotron DESY, D-22607 Hamburg, Germany}\\[0.1cm]
{\it $^2$ Petersburg Nuclear Physics Institute, Gatchina 188300, St. Petersburg, Russia}\\[0.1cm]
{\it $^3$ School of Physics and Astronomy, University of Southampton,\\Highfield, Southampton SO17 1BJ, UK}\\[0.1cm]
 \end{center}

\vspace*{3 cm}

\begin{center}
{\bf Abstract} \end{center}

We analyze, in leading and next to leading order of the BFKL equation,
the effects of the quantization of the singularities of the $j$-plane, t-channel
partial waves due to the imposition of appropriate infrared and ultraviolet boundary conditions.
We show that the intercepts, $\omega_n$ of the Regge poles, which contribute significantly to the
gluon density in the kinematic region measured at HERA and which can be calculated in QCD and in a supersymmetric
extension of QCD, are substantially modified by Beyond the Standard Model (BSM) effects. We also develop
a physically motivated heuristic model for the infrared boundary condition and apply it to the gluon
density. We argue that, using this type of model, the analysis of present and future low-$x$ data could 
allow one to detect supersymmetry at a high energy scale.

\vspace*{3 cm}

\begin{flushleft}
\end{flushleft}

\newpage

\section{Introduction}
The BFKL equation determines the high energy behaviour of the virtual gluon-gluon scattering amplitude in Regge limit, in which the cms energy, $\sqrt{s}$, is much larger than the transverse momenta, $k,k'$, of the gluons. It was derived in the fixed coupling constant case by resumming  {\it all} the Feynman diagrams describing gluon-gluon scattering in the leading or next-to-leading order.   The scattering amplitude displays a scale invariance such that it can be described  solely by functions of  ratios of  transverse momenta $k$ and $k'$. 

A common application  \cite{AMKS,KMS,CCS}
of the BFKL equation is to use it as an evolution equation in rapidity, $y$
and transverse momentum $k$ for some large-rapidiity amplitude, namely
 \beq \frac{\partial}{\partial y} {\cal A}(y,t) \ = \
  \int dt^\prime {\cal K}(\alphabar,t,t^\prime) {\cal A}(y,t^\prime)  \label{evolut1} \eeq 
where 
$$ t \ = \ \ln\left(\frac{k^2}{\Lambda_{QCD}^2}\right). $$
and  
the coupling $\alphabar$ runs with transverse momentum.
This application is often used as it
lends itself relatively easily to an extension of the DGLAP formalism to very low values of
Bjorken-$x$, where the pure DGLAP formalism is known to break down.

In order to solve this evolution equation, one requires as input the amplitude at some rapidity $y$
for all values of transverse momentum $t^\prime$.  One could naively expect that eq.~(\ref{evolut1}) could be directly solved for large $y$ and $t$ well in the perturbative region because it is known that the BFKL kernel $K(t,t')$ is quasilocal in $t$, i.e.  it diminishes when $|t-t'|$ is large. However, as was carefully investigated e.g. in ref.~\cite{AMKS}, the BFKL equation is  not only an evolution in $y$ but also in virtualities $t$, which leads to a substantial diffusion  into the low transverse momenta region where perturbative QCD cannot be valid.  Therefore, the authors of  ref.~\cite{AMKS} proposed a modification of the BFKL equation by imposing  a low and high cutoff in $t$ on the BFKL integral,  a procedure which is today widely accepted.   However, such a cutoff implies that the amplitude actually vanishes below a certain
transverse momentum, rather than becoming non-perturbative. In fact, due to the growth of the coupling
constant at small $k_T$, the amplitude could grow in this region and the vanishing of it
at the infrared cutoff looks unnatural.


In ref.~\cite{KLRW} we proposed to solve  eq.~(\ref{evolut1}) by the Green function method which does not require any cutoff on the BFKL integral. Instead,  we 
assumed that the non-perturbative infrared region of QCD
imposes a certain phase on the oscillatory parts of the eigenfunctions at some small transverse
momentum. 
This treatment of the infrared boundary leads to a discrete set of eigenvalues, $\omega_n$, of the BFKL kernel, since only certain values permit the
construction of eigenfunctions which simultaneously obey these phase conditions at low transverse momentum {\it and} the large transverse momentum
boundary conditions imposed by the asymptotic freedom.
 This is in contrast to the  ``usual'' treatment \cite{AMKS,KMS,CCS}, in which
a lower transverse momentum cutoff is imposed on the amplitude, i.e. the amplitude is assumed to vanish below the cutoff. 
In our approach,  the amplitudes are particularly sensitive to the exact values of the discrete $\omega_n$ which are related to the non-perturbative phases, $\eta_n$, at the cutoff. These phases are determined, in turn, by  the gluon-gluon interactions of the non-pertubative QCD, which lead to  rich structures below this cutoff.   

 A  rigid cutoff (either UV or IR) destroys
the scale invariance of the BFKL kernel - and hence the validity
of scale covariant solutions. As pointed out in \cite{AMKS}, the imposition of such 
cutoffs has no effect on the position of the leading singularity (the exponent of $x$ in structure functions)
but does affect the pre-factor, which is controlled by the form of this singularity.
 We find that the subleading singularities are also essential in order to obtain a good fit
to HERA data. In our application of BFKL dynamics,
 the scale and conformal invariance (which is central to the BFKL formalism
for fixed coupling)  is broken in a contolled way, namely only 
through the running of the coupling and we assume ``quasi-conformal'' solutions in which
the exponent of the transverse momentum  varies slowly in order to compensate for the
change in the coupling in an accordance with a generalized DGLAP dynamics..

In our previous paper \cite{KLRW} we have shown that HERA $F_2$ data, at low $x$, can be described
very well by the gluon density constructed from the discrete spectrum of eigenfunctions of the BFKL kernel.
 The spectrum contained many eigenfunctions, ${\cal O}(100)$, with eigenvalues $\omega_n$ varying from $\omega_1 \sim 0.25$ to $\omega_n \sim 0.5/n$ for large $n$.
 This  first successful confrontation of the BFKL formalism \cite{BFKL} with data led to the
 unexpected question as to whether the HERA data are sensitive to the Beyond Standard Model (BSM) effects. These effects, although only present at scales  that are much higher than the  region of HERA  data, can nevertheless affect the quality of the fits to data since BSM effects change the running of the coupling  and consequently also substantially change the values of $\omega_n$.


This seems somewhat counter-intuitive. 
One may ask how it can be possible
that a fit to data at relatively low energies can be sensitive to  corrections due
to loops of particles whose masses are far in excess of those energies.
The crucial point is that the above-mentioned large transverse momentum boundary conditions, imposed by the running
of the coupling,  can occur at very high $k_T$ scales \footnote{ Note that this means that formally we determine the eigenvalues   at asymptotically large initial energies.}.   For $\omega \, <  \sim \,  0.1 $ this is already above the scale at which one might expect
BSM physics to occur. The value of the discrete eigenvalues arise from an interplay between these ultraviolet boundary conditions and the infrared boundary condition arising from the imposition of an infrared phase. It is in this sense that in our treatment of the BFKL formalism
there is communication between high and low energy scales. We make the assumption that the allowed eigenvalues can be obtained from the
BFKL equation supplemented by an infrared phase condition in a {\it process-independent} way, i.e. without needing to impose any
infrared or ultraviolet cuts on the integration over transverse momentum. Once these eigenvalues are calculated their corresponding
eigenfunctions are  convoluted with the necessary impact factors in order to obtain the required
amplitudes. In this way, it turns out   that even though these amplitudes, in accordance with kinematical constraints,  never involve diffusion into transverse momenta above the
threshold for BSM physics,  the rapidity dependence of such amplitudes is affected by the substantial changes in the eigenvalues arising as a result
of BSM physics.

To understand how the running of the coupling constant can have such far-reaching consequences  we derive analytically (in Section 2)
 the main properties of the discrete pomeron solution using the LO BFKL equation.   This derivation provides 
 a  qualitative physical 
 explanation of the  mechanism by which the BSM effects modify  the discrete pomeron structures and lead to a genuine change of the eigenvalues and eigenfunctions. It  also elucidates  the 
  role of the infrared phases which define the boundary condition and which can be {\it indirectly determined} from data.  This explanation
 is then carried over into the NLO evaluation, which was used for data analysis and was performed 
 numerically. 
 

As a popular example of BSM effects we have chosen the N=1 supersymmetry and modified the 
$\beta$-function and the kernel  of the BFKL equation to include the contributions from the  superpartners.
We then describe the full NLO evaluation of the Discrete BFKL Pomerons (DP) with collinear resummation \cite{salam}.
This allows us to show that the eigenvalues, $\omega_n$, at larger $n$, have a genuine sensitivity to BSM physics because the support of the corresponding eigenfunctions extends to very high virtualities.   The values of $\omega_n$  are determined (to large extent)  by the running  of $\alpha_s$ and the  properties of the BFKL kernel in the high virtuality regions, where BSM effects dominate and QCD NLO correction are very small. This is also the reason why these eigenvalues are not sensitive to a particular choice  of the infrared boundary. All these properties  are discussed in detail in Section 2.

In Section 3 we then show  that it is possible to construct a
physically self-consistent infrared boundary condition 
which determines the properties of the gluon density to be in agreement with data. Finally we   confront the DP gluon density 
  with the HERA $F_2$ data and show that within our model for the infrared boundary condition we obtain 
indirect
 evidence of a supersymmetric threshold in a multi TeV range. 


In Section 4 we discuss our results with particular emphasis on the 
role of the universal Green function and the momentum
conservation.   We also discuss the dependence of our fits on the choice of the infrared boundary condition.  
  Section 5 presents a summary.

\section{The Discrete BFKL Pomeron}

The forward amplitude for a diffractive process with rapidity (or rapidiy gap) $y$
 is determined by the QCD pomeron and may be written
\beq {\cal A}(y) \ = \  \int d\omega  \int dt \int dt^\prime \Phi_u(t)  \Phi_d(t^\prime) e^{\omega y}
 \tilde{{\cal G}}_\omega(t,t^\prime),  \label{unintglue} \eeq
where  $ \tilde{{\cal G}}_\omega(t,t^\prime)$ is the Mellin transform of a universal
 (i.e. process independent) Green function
 ${\cal G}(t,t^\prime, y-y^\prime)$,
 \beq {\cal G}(t,t^\prime, y-y^\prime) \ = \ \int d\omega \tilde{{\cal G}}_\omega(t,t^\prime)
 e^{\omega(y-y^\prime)}  \label{mellgreen}. \eeq
The process dependence enters only through the impact factors, $\Phi_u, \, \Phi_d$ 
at the top and bottom of the gluon ladder which depend on the transverse momenta of
the gluons and may also depend on other kinematic variables.
The integral over $\omega$ in the inverse Mellin transform, (\ref{mellgreen}), is performed over a contour parallel to the imaginary axis,  to the right
of all singularities of the Green function.
 This Green function obeys the equation 
\beq \omega \tilde{{\cal G}}_\omega(t,t^\prime) - \int dt^{\prime \prime} {\cal K}(\alphabar,t,t^{\prime\prime})
  \tilde{{\cal G}}_\omega(t^{\prime\prime},t^\prime) \ = \ \delta(t-t^\prime), \eeq
which is solved by determining the set of eigenfunctions of the BFKL kernel, subject to certain boundary conditions.
If we allow the coupling $\alphabar$ to run with $t,t^\prime$, which in LO means replacing it
by $\sqrt{\alphabar(t)\alphabar(t^\prime)}$ then, as we explain below, the UV boundary condition
 - namely that the eigenfunctions decay as $t \to \infty$ is  automatically implemented. The infrared
boundary condition is imposed by requiring that the eigenfunctions have some given non-perturbative phases
at some low value of $t$ \footnote{ These phases should be universal (process independent).}.  As was first shown in ref.\cite{lipatov86} and we explain again below, the combination
of the UV boundary condition (which determines the oscillation phase at $t=t_c$, where the oscillatory
behaviour changes to an exponentially decaying behaviour, compatible with a DGLAP analysis
in the double logarithmic limit) and the infrared phase condition leads  (for positive $\omega$)
to a discrete set of allowed
eigenvalues $\omega_n$ with their attendant eigenfunctions, so that 
the Green function may be written as
\beq   \tilde{{\cal G}}_\omega(t,t^\prime) \ = \  \sum_n 
 \frac{f^*_{\omega_n}(t^\prime) f_{\omega_n}(t)}{\omega-\omega_n}
 \ + \ 
 \frac{1}{2\pi i}\int_{-\infty}^0 d\omega^\prime 
\frac{ f^*_{\omega^\prime}(t^\prime) f_{\omega^\prime}(t)}{\omega-\omega^\prime+i\epsilon}\label{nom} \eeq

No further cuts from kinematic constraints on $t$ are imposed in the determination
of the eigenfuntions and eigenvalues. However, as we discuss in detail
 in section 4, owing to the quasi-local nature
of the BFKL kernel ${\cal K}$,  this Green function is
rapidly attenuated for large $|t-t^\prime|$, which means that when it is 
 inserted into the expression
(\ref{unintglue}) in order to obtain the unintegrated gluon density, the diffusion
into values of $t$ substantially above the region of support of the impact factor,
$\Phi_p(t)$ is highly suppressed, thereby automatically  limiting
 the gluon virtuality to be small compared with the incoming energy.

The infrared non-perturbative phases, $\eta_n$, which determine the values of the discrete eigenvalues, $\omega_n$,
are in general $\omega$ dependent, but they must lie within a range of magnitude $\pi$,
so that the quantum number $n$ represents the number of oscillations  of the
eigenfunction  between the scale, $t_0$, at which the infrared phase condition
is applied and the ultra-violet scale, $t_c$, at which the oscillatory behaviour
becomes an exponentially decaying one. 
As pointed out in ref.~\cite{KLRW}, an $\omega$-dependent
infrared phase  condition is necessary in order to be able to express an impact factor
with support only for small $t$ in terms of the discrete eigenfunctions, since the
frequency of oscillation of these eigenfunctions at small $t$ is always below $\sim \, 0.7$. In fact, after imposing this non-perturbative phase, the eigenfunctions  form an almost complete set of functions  in the region of comparatively small $k \sim 1-10$ GeV.

The existence of this set of discrete eigenfunctions is consistent with the 
known fact
that in the Regge regime, the amplitude is determined by a set of Regge poles.
The imposition of infrared phases does not in any way violate the kinematical constraint,
but the ensuing discrete spectrum of eigenvalues has a very significant effect
on the fitting of the results of this modified BFKL formalism to HERA data on the
structure functions at low-$x$.

The continuum contribution (for negative $\omega$)
 is not significant for sufficiently small values of $x$ at any given $t$.
However, for a given $x$, as $t$ increases these contributions become more
significant and are essential in order for this formalism to match the
double logarithmic  limit of the DGLAP approach for sufficiently large $t$.

The value of $t_c$, at which the oscillatory behaviour converts into an exponentially
decaying one, increases linearly with eigenvalue number $n$. For $n \, \gtrapprox \, 3$
this occurs at values of $t_c$ above the scale at which one may expect to see physics
beyond the standard model. The running of the coupling is therefore affected
by the presence of thresholds for such new physics and this in turn affects the positions
of the discrete eigenvalues $\omega_n$. For sufficiently low values of $x$ the contribution
from all but the first two or three eigenfunctions is negligible. However, we find that
for values of $x$ which are probed at HERA there is a sizable contribution
from these higher eigenfunctions.  This means that even though the transverse momenta do {\it not}
diffuse into  regions of $t$ where the particles of  new physics are actually
produced, the shifts in the positions of the eigenvalues due to  
new physics affects the $x$-dependence of the unintegrated gluon density. Despite the
fact that these effects are small, the high quality of the HERA data means that the
quality of the fit  is significantly affected by the possibility of
new physics at high energies.

We now show how this works in detail.

\subsection{LO evaluation}

We begin this section 
by reviewing  the argument
of \cite{lipatov86} which led to a modification of the BFKL formalism
 which gives rise to discrete poles rather than a cut in the $\omega$-plane of the $t$ channel partial waves.

We consider the case of the leading order BFKL equation \cite{BFKL}
with running coupling also taken to leading order so that (for $t \, > \, 0$)
\beq \alphabar \ \equiv \frac{C_A \alpha_s}{\pi} 
 \ = \ \frac{1}{\betabar t} \label{beta} \eeq
where
\beq \betabar \ \equiv \  \frac{\beta_0}{4 C_A} \ = \ \frac{11}{12} - \frac{n_f}{18}. \eeq
The Hermitian BFKL kernel may be written as
 $$ \sqrt{\alphabar(t) \alphabar(t^\prime)} {\cal K}_0(t,t^\prime), $$
where
\beq \int dt^\prime {\cal K}_0(t,t^\prime) e^{i\nu t^\prime} \ = \ 
 \chi_0(\nu) e^{i\nu t} \label{bfkl1} \eeq
\beq \chi_0(\nu) = 2 \Psi(1) - \Psi\left(\frac{1}{2}+i\nu\right)
 - \Psi\left(\frac{1}{2}-i\nu\right) \eeq
Note that the parameter $\nu$ may be real or imaginary for real eigenvalues $\chi_0(\nu)$.

The eigenfunctions, $g_\omega(t)$  of this Hermitian kernel obey the eigenvalue equation
\beq \int  dt^\prime  \sqrt{\alphabar(t) \alphabar(t^\prime)}  {\cal K}_0(t,t^\prime)
 g_\omega(t^\prime) \ = \ \omega g_\omega(t) \eeq
These eigenfunctions form a complete orthonormal set
\beq \int dt g_\omega(t)g^*_{\omega^\prime}(t) \ = \  2\pi\delta(\omega-\omega^\prime)
 \label{norm}, \eeq
  assuming (for the moment) a continuous spectrum for the eigenvalues, $\omega$. 
  
They can be  obtained by defining a set of functions $f_\omega(t)$:
 \beq f_\omega(t) \ = \ \frac{g_\omega(t)}{\sqrt{t}} \eeq
which obey the eigenvalue equation 
\beq  \alphabar(t) \int  dt^\prime    {\cal K}_0(t,t^\prime)
 f_\omega(t^\prime) \ = \ \omega f_\omega(t) \eeq

Using eq.(\ref{beta}) we have
\beq \int dt^\prime  {\cal K}_0(t,t^\prime) f_\omega(t^\prime)
 \ = \ \betabar \omega t f_\omega(t)  \label{bfkl2} \eeq

Taking the Fourier transform
 \beq f_\omega(t) \ = \ \int d\nu e^{i\nu t} \tilde{f}_\omega(\nu) \label{ftrans} \eeq
and using eq.(\ref{bfkl1}) we have a first-order differential equation
\beq \frac{d}{d\nu} \tilde{f}_\omega(\nu)  \ = \ -\frac{i}{\betabar\omega} \chi_0(\nu) \tilde{f}_\omega(\nu)
\eeq 
which has a well known solution 
\beq \tilde{f}_\omega(\nu)
 \ = \ \exp\left(-\frac{i}{\betabar\omega}\int^\nu \chi_0(\nu^\prime)d\nu^\prime\right) 
 \ = \ \factor^{1/(\betabar \omega)},  \label{sol0} \eeq
 In this way we obtain 
\beq f_\omega(t) \ = \ \frac{1}{\sqrt{2\pi  \omega}} \int_{-\infty}^{+\infty} d\nu e^{i\nu t}  
\factor^{1/(\betabar \omega)},  \label{sol} \eeq
where the pre-factor is taken such that the normalization condition (\ref{norm}) is obeyed.

The integral over $\nu$ can be performed numerically over a suitable contour. A very good 
approximation to this integral (for small $\omega$) is obtained by the saddle-point approximation,
(equivalent to solving eq.(\ref{bfkl2}) using the semi-classical approximation).
The saddle point, which is a function of $t$, $\nu_s(t)$,  is obtained from the solution to
 \beq \chi_0\left(\nu_s(t)\right) \ = \ \betabar \omega t \label{saddle}\, . \eeq 
We consider two regions depending on whether $t$ is greater or less than a critical
point, $t_c$,  given by
 \beq \betabar \omega t_c \ = \ \chi_0(0) \ = \ 4 \ln 2 \label{crit}\, . \eeq

\begin{figure}
\centerline{\epsfig{file=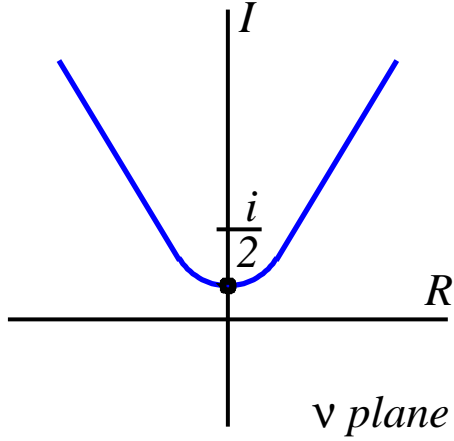, width = 8 cm}}
\caption{ Integration contour (blue line) on the $\nu$ plane for  $t> t_c$. The black dot shows the position of the saddle point, $\nu_s$. \label{contour1} }
\end{figure}

\noindent $t \ > t_c:$ \\
In this case there is a single saddle-point on the positive imaginary 
axis, shown in Fig.~\ref{contour1}.

\noindent If we define $\gamma_s$ by
\beq \gamma_s \ = \ \frac{1}{2}+i\nu_s \eeq then at the saddle-point, $\gamma_s$,
 is the solution to
\beq  \chi_0(\gamma_s) \ \equiv \ 
2\Psi(1) - \Psi(\gamma_s) - \Psi(1-\gamma_s) \ = \ \betabar \omega t\, \label{ll2} \eeq
 $\gamma_s$ is in the range
 $$ 0 \ < \ \gamma_s \ < \ \frac{1}{2} $$
The contour of integration is deformed so that it becomes the contour of steepest descent
obtained from the solution to
 $$ \arg \left\{   
 \int_{\nu_s}^\nu \chi(\nu^\prime) d \nu^\prime - \chi(\nu_s) \left(\nu-\nu_s \right)  
 \right\} \ = \ -\frac{\pi}{2}  $$
Near the saddle point the contour runs parallel to the real axis  
but for very large $|\nu|$ it runs parallel
to the imaginary axis. In the saddle-point approximation, we obtain (using eq.(\ref{ll2}))
\beq  f_\omega(t) \ = \ \sqrt{\frac{1}{2\chi_0^\prime(\gamma_s)}} e^{-t/2} e^{\gamma_s t} 
 \left[ e^{(\gamma_s-1/2)\Psi(1)} \frac{\Gamma(\gamma_s)}{\Gamma(1-\gamma_s)}
 \right]^{1/(\betabar \omega)} \eeq
This is an exponentially decreasing function of $t$. Moreover $\gamma_s$
can be related to the anomalous dimension in the DGLAP formalism since
\beq \frac{d}{dt} \left(  e^{t/2} f_\omega(t) \right) \ = \ \gamma_s
 \left(  e^{t/2} f_\omega(t) \right). \eeq
 From eq.(\ref{saddle}), the anomalous dimension is
 $$ \gamma_s \ \approx \, \frac{\alphabar(t)}{\omega} +O\left( \frac{\alphabar(t)^2}{\omega^2}\right)$$
  in agreement with DGLAP for small $\gamma_s$.
 
 \begin{figure}
\centerline{\epsfig{file=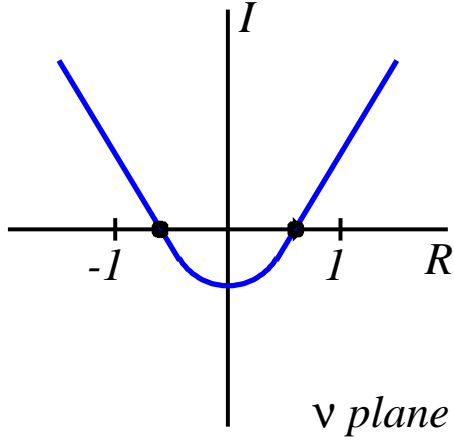, width = 8 cm}}
\caption{ Integration contour (blue line) on the $\nu$ plane for  $t<t_c$. The black dots show the positions of the saddle points, $\pm |\nu_s|$. \label{contour2} }
\end{figure}

\noindent $t \ < t_c  :$ \\
Here we have two saddle points lying on the real axis at $\pm \nu_s$, shown in Fig.~\ref{contour2}. The positions of the saddle points are obtained from
 \beq 2\Psi(1)  - 2 \Re e \left\{  \Psi\left(\frac{1}{2}+i\nu_s\right) 
\right\} \ = \ \betabar \omega t. \eeq
We need to integrate around both of these saddle points, taking
a contour of steepest descent in the vicinity of the saddle-points,
 which in this case is inclined at an angle
of $\, \pm \pi/4$ to the real axis and enclose  the positive imaginary axis at large $\nu$. The saddle-point
approximation then yields
 \beq f_\omega(t) \ = \ \sqrt{\frac{2}{\chi^\prime(\nu_s)}}
 \sin \left(\nu_s t + \frac{\phi(\nu_s)}{\betabar\omega}+\frac{\pi}{4}  \right),
\label{pi4} \eeq
where
 \beq \phi(\nu_s) \ = \ \mathrm{Arg} \left\{ e^{-2i\Psi(1)\nu_s}
 \frac{\Gamma(1/2+i\nu_s)}{\Gamma(1/2-i\nu_s)} 
  \right\} \eeq
The inclusion of $\pi/4$ in the phase in eq.(\ref{pi4}) ensures a matching of the 
solutions at $t=t_c$. Near $t=t_c$ the solution is given by  an Airy function.

For $t<t_c$, we have an oscillatory solution which does {\it not} lend itself to
a match to the DGLAP formalism - in this regime a DGLAP analysis is {\it not} appropriate,
since in this region the saddle-point $\gamma_s$ is complex and double-valued, i.e.
 $$ \gamma_s \ = \ \frac{1}{2} \pm i |\nu_s|, $$
and cannot be related to the (real) anomalous dimension of the DGLAP formalism. In this region the DGLAP equation is not valid and the BFKL equation can be considered as a  generalized (quantized) version of the DGLAP equation. 

Recall that the saddle-point, $\nu_s(t)$, is a function of $t$ and so we do not have constant
frequency oscillations. As $t \, \to \, 0$, $\nu_s$ tends to a value $\nu_0 \, \approx 0.635$
and we have constant frequency oscillations in the infrared limit. As $t$ increases this frequency
decreases, becoming zero at $t=t_c$.

The infrared phase at $t=0$, calculated from perturbative QCD (with $ t \, > \, 0$), is then given by
\beq \eta_0 \pi  \ = \  \frac{\phi(\nu_0)}{\betabar \omega}+\frac{\pi}{4}, \eeq
($\phi(\nu_0) \, \approx \, 0.96$).
This phase is only determined up to a multiple of $\pi$. We now make a very general assumption  that the infrared 
properties of QCD fixes  this phase (in general as a function of $\omega$) to be $\eta(\omega)$,
where the function $\eta(\omega)$ is determined from the non-perturbative regime of QCD ($t \, \le \, 0$).
The matching of the two phases $\eta_0$ and $\eta(\omega)$ in the semi-classical solution, 
eq.(\ref{pi4}),
then restricts the allowed values of $\omega$ to a discrete set $\omega_n$ that satisfy the
equation
 \beq \frac{\phi(\nu_0)}{\pi \betabar \omega_n}  \ = \ \eta(\omega_n) + \left( n-\frac{1}{4}\right) 
 \ \ \ \, (n \, = \, 1,2,3 \cdots )
 \label{omegan-lo}
  \eeq 
The function $\eta(\omega)$ could be a constant
(as originally proposed in \cite{lipatov86})
but in general it can vary with $\omega$. Because of periodicity it can take values  in the interval  between
 $0.25$ and $-0.75$ only.  
 Although $\eta(\omega)$  cannot be determined from the perturbative analysis described here,
 its restricted range limits its   effect on the determination of the eigenvalues
(see Section 2.5). 
However, its variation with $\omega$ is very important in the construction of the gluon density, 
 (see Section 3).

The above analysis shows clearly that the solution of the BFKL equation has to be given by the set of discrete eigenfunctions, whose support in the virtual gluon transverse momentum is determined by the critical point, $t_c$, eq.~(\ref{crit}), and whose phase, $\eta(\omega)$,
 at some low transverse momentum is determined by the non-perturbative sector of QCD.
Effectively the boundary conditions modify the BFKL kernel so that it may be written as
 \beq {\cal K}(t,t^\prime) \ = \ \sum_n \omega_n f_n^*(t) f_n(t^\prime) \label{kernel3} \eeq
This coincides with the (LO) kernel
$$ \alphabar(t) K_0(t,t^\prime) $$
provided it acts on a function, $f(t)$ which may be written as a superposition of
the eigenfunctions $f_n(t)$:
$$ f(t) \ = \ \sum_n a_n f_n(t). $$
In this way we have supplemented the kernel with both infrared and ultraviolet boundary
conditions. The infrared boundary conditions arise from the non-perturbative sector of QCD,
but the ultraviolet boundary conditions arise naturally from the asymptotic freedom of QCD.
Importantly, the behaviour of the eigenfunctions in the ultraviolet is controlled by a
critical value, $t_c$, of transverse momentum, which grows almost linearly with $n$ in accordance with the fact that the period of oscillations is practically  independent on $n$.
 Therefore, the $n$-dependent boundary condition leads to qualitatively different results
from those obtained using a kernel in which boundary conditions are effected simply by a cutoff on a wave function.

Furthermore we note that the value of this critical transverse momentum depends almost entirely
 on the eigenvalue, $\omega$,
which decreases like $1/n$, for large $n$, as the quantum number $n$ increases,
(see eq.~(\ref{omegan-lo}).
 This means that in turn the value of the critical transverse momenta, $k_c$,
   increases exponentially as $\omega$ decreases, so that the $n^{th}$ critical momentum is given by 
 (inserting eq.(\ref{omegan-lo}) into eq.(\ref{crit}))
\beq k_c^{(n)} \ = \ \Lambda_{QCD} \exp\left\{ \frac{2 \pi \ln 2}{\phi(\nu_0)} \left(
   n-\frac{1}{4}+\eta(\omega_n)  \right) \right\}  \ \approx \  \ \Lambda_{QCD} \, e^{4.5\,  n} 
 \label{lokcrit}
 \eeq

Finally let us note that our solution of the BFKL equation is similar to the  WKB method for the bound-state 
solution of the Schr\"{o}dinger equation in the semi-classical approximation;  the critical point, $t_c$, is analogous to the turning point $x_c$ where the potential is equal to the energy.  Inside a potential well the solutions are oscillatory and outside they 
decay exponentially. 
 This shows that the solution of the  BFKL equation (\ref{evolut1}) consists of the superposition of the    bound state eigenfunctions of the two gluon system with pseudo-energies given by the eigenvalues $\omega_n$. 
 Knowledge of the eigenvalues gives important information about the interactions between gluons, both
in the infrared and ultraviolet regions of $k$.
We note that  in the solution of the BFKL equation,  the oscillations  of the eigenfunctions at large $k$  should cancel each other  in accordance with the kinematical constraints provided by the beam energy of the experiment. This imposes additional restrictions on the non-perturbative phases $\eta(\omega)$, see below. 


\subsection{Threshold effects}
The above analysis assumes that $\betabar$ is a constant
so that the coupling $\alphabar(t)$ is given simply
by eq.(\ref{beta}). However, we know that there are
thresholds at $t=t_i$ where heavy
 flavour quarks can be produced, and also there may be extra thresholds
arising from BSM physics with a threshold
 ( that according to eq.(\ref{lokcrit}) can be large)
 below
$t=t_c$. This means that eq.(\ref{sol}) can only be used as a solution for $f_\omega(t)$ between thresholds. As an example, suppose that there
is only one threshold, at $t=t_t$  below  the critical point,
$t_c$, and that $\betabar$ takes the value $\betabar^>$ above this threshold
 and $\betabar^<$ below. At $t\ge t_t$ we have
\beq f_\omega(t) \ = \ \frac{1}{\sqrt{2\pi \betabar \omega}}
 \int d\nu\, e^{i\nu t} \factor^{1/(\betabar^>\omega)} \label{thres1} \eeq
and for $t < t_t$  we have
\beq f_\omega(t) \ = \ A
 \int d\nu \, e^{i\nu t} \factor^{1/(\betabar^<\omega)} f_\omega(t_t)\label{thres2} \eeq
with the constant $A$ chosen to be
\beq A^{-1} \ = \ \int d\nu \, e^{i\nu t_t}\factor^{1/(\betabar^<\omega)}, \eeq
so that the solutions match at $t=t_t$.

In the saddle-point approximation, we can handle such thresholds by noting
that eq.(\ref{sol}) can be written as
\beq f_\omega(t) \ = \ \frac{1}{\sqrt{2\pi\omega}}
 \int d\nu e^{iS(\nu,t)/\omega} ,  \label{semi} \eeq
where the  ``action''  $S(\nu,t)$ is given by
\beq S(\nu,t) = \omega \nu t - \frac{1}{\betabar} \int_0^\nu \chi_0(\nu^\prime) d\nu^\prime . \eeq
At the saddle-point, $\nu=\nu_s(t)$, upon integrating by parts this may be rewritten
as
 \beq S(t) \ = \ \omega \int_{t_c}^t \nu_s(t^\prime) dt^\prime \label{phase2} \eeq
where the function $\nu_s(t)$ is given by eq.(\ref{saddle}). (We have used the notation
 $S(t)$ to denote $S(\nu_s(t),t)$ - it is now a function of $t$ only).
Here we see explicitly that the saddle-point approximation for integral~(\ref{semi}) is equivalent to the semi-classical 
approximation. Replacing the integral over $\nu$ in eq.(\ref{semi}) 
by the value of the integrand at the saddle-point
we obtain a solution which obeys the differential equation
\beq \frac{d}{dt} f_\omega(t) \ = \ \chi^{-1}(\betabar \omega t) f_\omega(t) 
 \label{semi2}  \eeq
The semi-classical approximation consists of the assumption that the solution to
eq.(\ref{bfkl2}) is the solution to eq.(\ref{semi2}) multiplied by a slowly varying factor,
which turns out to be the same as that obtained in the Gaussian integral around
 the saddle-point in  eq.(\ref{semi}).   
In analogy with the WKB approximation in the Schr\"{o}dinger equation, there
exists a critical point, $t_c$, at which  the  approximate solution  
changes from an oscillatory function to an exponentially decaying one. The assumption of a 
slowly varying pre-factor breaks down at this point, but the solutions either side of the critical
point can be matched using a suitable Airy function. It is this matching, together with some
property of the behaviour at $t=0$  that determines the allowed eigenvalues.

Thresholds are handled in general by replacing eq.(\ref{saddle}) by the more general
relation
 \beq \chi_0\left(\nu_s(t)\right) \ = \ \frac{\omega}{\alphabar(t)}.  \label{saddle2} \eeq
$\alphabar(t)$ may now be determined using the $\beta$-function with appropriate thresholds.

The eigenfunctions for $t \, \ll \, t_c$ now take the form
\beq f_\omega(t) \ = \ \frac{C}{\sqrt{\chi_0^\prime(\nu_s(t))}}
 \sin\left(\frac{S(t)}{\omega}+\frac{\pi}{4}\right) \label{ef2} \eeq
and the semiclassical quantization condition on the allowed eigenvalues becomes
 \beq \frac{S(0)}{\omega_n} \ = \ \left(
\eta(\omega_n) + n-\frac{1}{4} \right)\pi 
 \label{spectrum} \eeq  
Comparing this with eq.(\ref{omegan-lo}) we see that $S(0)$ 
is independent of $\omega$ \footnote{For sufficiently  large $n$ we can see this
directly  since
 $S(0)\, \sim \, {\cal O}(\omega_n t_c^{(n)})$ and whereas $t_c^{(n)} \, \propto \, n$,
 $\omega_n \, \propto \, 1/n$.}.
  However, its value
 clearly depends on the positions of the thresholds in $\betabar$, eqs. (\ref{thres1}) and (\ref{thres2}).
Already for $n \ge 3$ the value of $k_c$ given by eq.( \ref{lokcrit})  exceeds many tens of TeV 
and this means that the spectrum is sensitive to any BSM physics.
The BSM effect changes the perturbative phase $\eta$ if the corresponding threshold, $k_t$, is below the critical point, $k_c^{(n)}$.  The effect of such a threshold can be readily estimated;
let us assume that for a given eigenvalue, $\omega$, the threshold 
$t_t \, \equiv \,  \ln(k_t^2/\Lambda_{QCD}^2) $ is the largest threshold, below the critical
point $t_c$, and that in the range $$ t_t \ < \ t \ < t_c,$$
$\betabar$ takes the value $\betabar^>$.
The exact solution for the LO BFKL with running coupling at $t=t_t$ is
\beq f(t_t) \ \propto \ \int d\nu e^{i\phi(\nu,t_t)}, \eeq
where
\beq \phi(\nu,t_t) \ = \ \nu \left(t_t-2\frac{\Psi(1)}{\betabar^> \omega}\right)
    + \frac{2}{\betabar^>\omega}\arg \left\{\Gamma\left(\frac{1}{2}+i\nu\right) \right\} \eeq    
We evaluate this integral  using the saddle-point method and assuming that  
the threshold occurs sufficiently close to the
critical point $t_c$ that $\nu_s$ is sufficiently small for the diffusion approximation to be
valid.
 The phase difference between the case where there is a threshold at $t=t_t$
and the case where there is no threshold is then
\beq \Delta \phi \ = \ \frac{2}{3  \omega \sqrt{14\zeta(3)}} 
\left[ \frac{1}{ \betabar^<} \left(4\ln 2 - \betabar^<\omega t_t \right)^{3/2}  -\frac{1}{ \betabar^>}  
\left(4\ln 2 - \betabar^> \omega t_t \right)^{3/2} \right] 
\label{deltaphi}
\eeq
where  $\betabar^<$ denotes the value of  $\betabar$ below 
 the threshold.

We observe that the above phase difference is substantial for $\omega \le 0.1$ and that this difference is not suppressed by the scale of the BSM physics. 
In this sense our analysis differs fundamentally from the treatment of the Standard Model as a low-energy
effective theory way below the thresholds of new physics. In the latter case,
logarithmic corrections can always be absorbed into unphysical renormalization constants 
of the renormalizable operators of the effective theory, leaving only
higher dimension operators whose coefficients are suppressed by powers of
the new-physics mass scale.

\subsection{NLO evaluation}

We have shown in~\cite{KLRW}  that the  BFKL integral equation  can be cast in the form of the pseudo-differential equation
\beq 
  \alphabar(t) \int  dt^\prime    {\cal K}_0(t,t^\prime)
 f_\omega(t^\prime) \ = \ 
\chi\left( -i \frac{d}{dt}, \alpha_s(t) \right) f_\omega(t) \ = \ \omega
f_\omega(t) \label{bfkl-lin}. \eeq 
Making the simplifying assumption that $\alphabar$ is given by eq. (\ref{beta})  and taking into
 account the collinear resummation \cite{salam} we can write the BFKL equation in the next-to-leading order
as
 \beq   \betabar \omega t f_\omega(t) = \left[\tilde{\chi}_0(\hat{\nu},\omega) 
+ \frac{1}{t\betabar} \xi(\hat{\nu})
 \right] \cdot f_\omega(t), 
 \label{next-saddle}\,  \eeq 
where the operator $\hat{\nu} = -i d/dt$,
$$  \tilde{\chi}_0(\nu,\omega) =2\Psi(1)-\Psi\left(\frac{1}{2}+i\nu+\frac{\omega}{2}\right)
 -\Psi\left(\frac{1}{2}-i\nu+\frac{\omega}{2}\right)  $$  
 and
 $$  \xi(\nu) =\chi_1(\nu) + \frac{1}{2}
 \left[2\Psi(1)-\Psi\left(\frac{1}{2}+i\nu\right) -  \Psi\left(\frac{1}{2}-i\nu\right)
 \right] \left[\Psi^\prime \left(\frac{1}{2}+i\nu\right) +  \Psi^\prime
\left(\frac{1}{2}-i\nu\right)\right]\, ,  $$
where $\chi_1$ denotes the NLO characteristic function~\cite{FL}. The function $\xi(\nu)$ does not have the poles of the third order of the form $\sim \, 1/(1/2\pm i\nu)^3$, in agreement the a
renormlaization group analysis \cite{FL}.

The equation (\ref{next-saddle}) can be considered as a quadratic equation in $t$
\beq
\left[\omega(t\betabar)^2-t\betabar\tilde{\chi}_0(\hat{\nu},\omega)-\xi(\hat{\nu})\right] \cdot f_\omega(t) = 0.
\label{quadrat-eq}
\eeq
We can convert this into a second order differential equation for the Fourier transform $\tilde{f}_\omega(\nu)$,
where $t$ is replaced by the operator $\hat{t}=i\frac{d}{d\nu}$. In the semi-classical approximation
in which  $\ln(\chi_0)$ and $\ln(\xi)$ are treated as slowly varying functions of $\nu$,
so that    
$$ \hat{t}^2\cdot \tilde{f}_\omega \approx \left(\hat{t}\cdot (\ln \tilde{f}_\omega)\right)^2 \tilde{f}_\omega,$$ 
this  second-order differential equation may be written as
\beq
\left[i\betabar \hat{t}-X^-(\nu,\omega)\right] \cdot
\left[ i\betabar \hat{t}-X^+(\nu,\omega)\right] \cdot \tilde{f}_\omega(\nu)=0
\eeq
where
$$ X^\pm(\nu,\omega) = \frac{1}{2\omega} \tilde{\chi}_0(\nu,\omega)  \pm \sqrt{\left(\frac{1}{2\omega}\right)^2 \tilde{\chi}_0(\nu,\omega) +\frac{1}{\omega}\xi(\nu) }, $$
with solution for $f_\omega(t)$ which is analogous to  the LO expressions of  eq. (\ref{ftrans}) and (\ref{sol0}):
\beq  f_\omega(t) = \int_{-\infty}^\infty d\nu \, e^{i\nu t}
 \exp \left(  -\frac{i}{\betabar}\int^\nu X^+ (\nu,\omega) d\nu   \right).  \label{neweq}  \eeq
For small $\omega$,  eq.(\ref{neweq})   may be approximated by
\beq f_\omega(t) \ = \ \frac{1}{\sqrt{2\pi  \omega}} \int_{-\infty}^{+\infty} d\nu e^{i\nu t}  
\factor^{1/(\betabar \omega)} e^{-\frac{i}{\betabar^2\omega t} \int^{\nu} d\nu' \xi(\nu') }.  \label{sol1} \eeq
Note that the saddle-point of the integral over $\nu$ in eq.(\ref{sol1}) occurs
at $\nu_s$ where the RHS of eq.(\ref{next-saddle}) vanishes as required for
 $\tilde{f}_\omega(\nu_s)$ to be a turning-point.

In a more general NLO approach, the BFKL equation   (\ref{bfkl-lin})  can be simplified using 
 the semi-classical approximation, i.e. assuming that the $t$-dependence of $\ln f_\omega(t)$ is large so that
\beq \left( \frac{d}{dt} \right)^r f_\omega(t) \ \approx \ 
 f_\omega(t) \left(\frac{ d \ln f_\omega(t)}{dt} \right)^r .\eeq 
The eq.~(\ref{bfkl-lin}) looks then like the non-linear differential equation
\beq \chi\left( -i \frac{d \ln f_\omega(t)}{dt}, \alpha_s(t) \right) \ =\ \chi(\nu(t),\alpha_s(t)) \ = \ \omega 
\label{nonlinear}. \eeq
As a result the frequency  $\nu(t)$ is a function of $t$ such that
  \beq 
  \omega \ = \  \left(\frac{\alpha_s(t) C_A}{\pi} \right)\chi_0(\nu) +
    \left( \frac{\alpha_s(t) C_A}{\pi} \right)^2 \chi_1(\nu) + \cdots
\label{hr}\eeq
The expression  (\ref{hr}) -including collinear resummation \cite{salam} - is the NLO analogue of the eq. (\ref{saddle2}).
The eq.(\ref{nonlinear}) has  a solution
 \beq f_\omega(t) \ = \ e^{i S(t)/\omega}  \label{eq232} \eeq
where
 \beq S(t) \ = \ \omega \int_{t_c}^t \nu(t^\prime) dt^\prime, \label{phase3} \eeq
The critical logarithmic transverse momentum, $t_c$, is the value
of $t$ for which $\nu(t)=0$. This condition is the NLO analog of eq.(\ref{crit}).
 For $t \, < \, t_c$, there are two real solutions
for $\nu(t)$ generating an oscillatory solution with a given phase, whereas for
$t \, > \, t_c$ the solution is on the positive imaginary axis, generating an
exponentially decaying function as $ t \, \to \, \infty$

Thus we see, that the solution for the eigenfunctions in semi-classical
approximation is analogous to that in leading order, but the function
$\nu(t)$ takes into account the NLO characteristic function as well as the
NLO running of the coupling and the threshold effects.
A further feature of threshold effects beyond leading order
is that it is not only the $\beta$-function that has steps at the thresholds but also
the NLO contributions to the characteristic function, $\delta \chi_1$, - corresponding to the
presence of new particles at some point in the ladders \cite{kotlip}. 

The semi-classical approximation is valid provided
 $$ \frac{ d\ln\left(\nu(t)\right)}{dt} \ \ll \ \nu(t) $$
This condition breaks down in the region $t \, \sim \, t_c$ where $|\nu(t)|$
is very small. However, in this region the eigenvalue equation approximates to
Airy's equation with solution
 \beq f_\omega(t) \ =  \  
 \mathrm{Ai}\left(\left(\frac{3}{2} \frac{S(t)}{\omega}\right)^{2/3} \right), \label{eq237} \eeq
For $ t \, \gg \, t_c$ the Airy function, $\mathrm{Ai}$, behaves as
\beq  \mathrm{Ai} \left(\left(\frac{3}{2} \frac{S(t)}{\omega}\right)^{2/3} \right)
 \ \sim  e^{-|S(t)|/\omega} \eeq
and for $ t \, \ll \, t_c$
\beq  \mathrm{Ai} \left(\left( \frac{3}{2} \frac{S(t)}{\omega}\right)^{2/3} \right) \
 \sim  \sin\left( \frac{S(t)}{\omega}+\frac{\pi}{4}\right) 
\label{eq238}\eeq
We therefore find that the solution eq.(\ref{eq237}) is a good approximation over the
entire range of $t$ and at the same time determines the phase of the oscillatory
 solution for $t \, = t_c$ required to match the oscillatory region and the exponentially
decaying region.
 As in the LO case, we make  a very general assumption that the infrared (non-perturbative) properties of QCD impose some phase, $\eta$, at $t=0$, defined up to an ambiguity of $n\pi$,
which can also depend on $\omega$. We find then that we can only match this phase to the perturbative one, determined from eq.(\ref{phase3}), for one value 
 of $\omega$ for each 
 integer $n$, where $n$ corresponds to the number of oscillations. 
 This leads to the quantization of the
 spectrum (i.e. discrete pomeron poles) given by eq.(\ref{spectrum}), in  keeping with the predictions of Regge theory.

In contrast to the LO evaluation,  in full NLO the eigenvalues and eigenfunctions can only be determined 
using numerical methods of iteration and integration. 
Their construction requires several steps. In the first step we determine the values of the frequency $\nu$
as a function of $\omega$ and $t$  from the  solutions of eq.(\ref{hr}).
  Then, the critical point, $t_c$, is determined as a  function of $\omega$
 from the condition $\nu(t_c)=0$. The phase function 
 $S(t)$, for a given $\omega$, is then found from eq.(\ref{phase3}).

In the next step the phase, $\eta$, at the infrared boundary has to be specified. In the leading order computation it was possible  to define it at $\Lambda_{QCD}$,  because the frequency $\nu$ is well defined at
 $t=0$, eq.(\ref{saddle}). For the NLO calculation, we obtain $\nu$ with the help of eq.(\ref{hr}),
 which is not valid at  $\Lambda_{QCD}$.
We therefore  defined it  as a phase condition at the lowest possible  value of the (logarithmic) 
 transverse momentum, $t=t_0$, which can be safely reached by the perturbative calculation (see also the discussion in Section 3.2).

\begin{figure}
\centerline{\epsfig{file=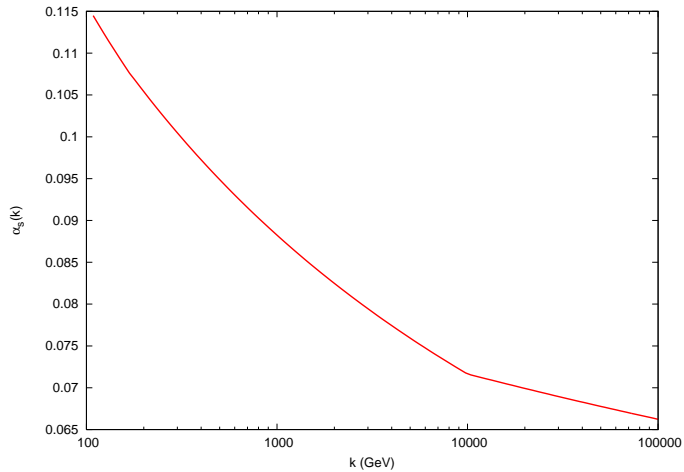, angle=270, width = 9 cm}}
\caption{ The running of $\alpha_s$ across a threshold for N=1 SUSY at 10 TeV \label{alphahigh} }
\end{figure}

\subsection{N=1 Supersymmetry at Various Thresholds}
We have chosen as example of  ``new physics''  the popular $N=1$ supersymmetric extension
of the Standard model above a given threshold in  $k_T$, which for simplicity we assume to be
a common mass threshold for all  superpartners.
 Below this threshold the running of the
coupling is governed by the $\beta$-function to two-loop order
\beq \beta_< \ = \ -\frac{\alpha_s^2}{4\pi} \left(\frac{11C_A}{3}-\frac{2}{3} n_f \right)
 - \frac{\alpha_s^3}{(4\pi)^2} \left(\frac{34 C_A^2}{3}+\left(\frac{10C_A}{3}+2C_F\right) n_f\right),
\eeq
where for the case of QCD, $C_A=3, \ C_F=4/3$ and $n_f$ is the number of active flavours.
Above the threshold, the $\beta$-function is given by
\beq \beta_> \ = \ -\frac{\alpha_s^2}{4\pi} \left(3C_A -  n_f \right)
 - \frac{\alpha_s^3}{(4\pi)^2} \left(6 C_A^2+\left(-\frac{2C_A}{3}+2C_F\right) n_f\right).
\eeq
This leads to a ``kink'' (discontinuity in the derivative) in the running of $\alpha_s$
at the threshold for N=1 SUSY, which can be seen in Fig.\ref{alphahigh}.

Furthermore, above the SUSY threshold, the NLO characteristic function, $\chi_1(\nu)$ acquires an additional contribution
\cite{kotlip} of
\beq \delta_f \chi_1(\nu) \ = \ \frac{\pi^2}{32} 
  \frac{\sinh(\pi \nu)}{\nu (1+\nu^2) \cosh^2(\pi \nu)}
   \left(  \frac{11}{4}+3\nu^2  \right) \label{dchif} \eeq
from the octet of Majorana fermions  (gluinos), and
\beq \delta_s \chi_1(\nu) \ = \ -\frac{\pi^2}{32} \frac{n_f}{C_A^3}
  \frac{\sinh(\pi \nu)}{\nu (1+\nu^2) \cosh^2(\pi \nu)}
   \left(  \frac{5}{4}+\nu^2  \right) \label{dchis} \eeq
from the squarks (in the fundamental representation).

\begin{figure}
\centerline{\epsfig{file=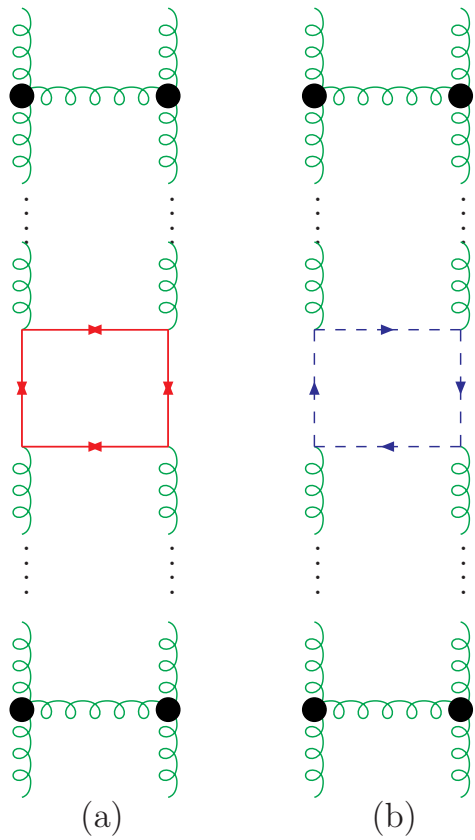, width=7cm }}
\caption{Typical graphs contributing to BFKL kernel involving (a) gluinos  or (b) squarks 
 \label{ladders}}
\end{figure} 
Typical graphs contributing to $\delta_s \chi_1(\nu)$ are shown in Fig. \ref{ladders}.
They contribute only at NLO level since the exchange of a fermion or scalar particle in the $t$-channel is suppressed in LLA
 \cite{FL}, and therefore  only contributes  at subleading logarithm order.

\subsection{ The Discrete Pomeron with and without SUSY}

\begin{figure}
\centerline{\epsfig{file=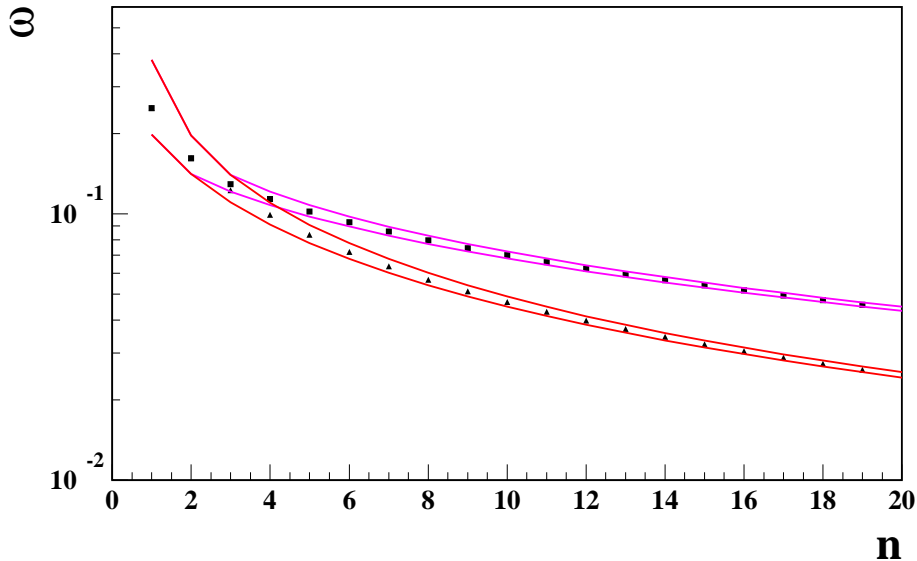, width=14cm}}
\caption{The eigenvalues  computed in the NLO evaluation  of the Standard Model (triangles)
 and SUSY at a threshold of 10 TeV (squares). The lines indicate the maximal possible spread  due to the  uncertainty of the phase ($\eta$)  choice.   \label{omegan}}
\end{figure}

\begin{figure}
\centerline{\epsfig{file=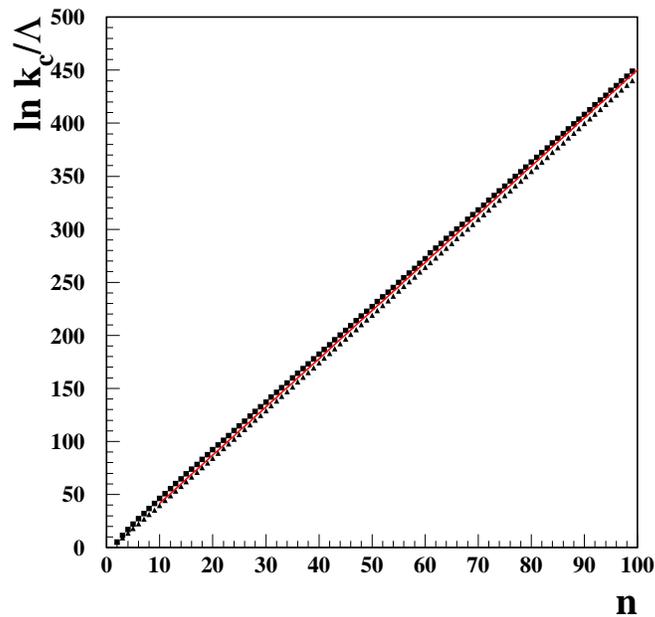, width=10cm}}
\caption{The critical point $k_c$  computed in the NLO evaluation  of the Standard Model (triangles)
 and SUSY at a threshold of 10 TeV (squares). The red lines show the LO computation
  which is not sensitive to any threshold effects.}   \label{kcrit}
\end{figure}

In this sub-section we investigate the properties of the Discrete BFKL Pomeron with
 and without SUSY contributions. For this example, we have  assumed that the SUSY threshold is at 10 TeV. 
  Fig.~\ref{omegan} shows  the spectrum of the eigenvalues $\omega_n$ computed in the NLO computation assuming that the non-perturbative phase,
 $\eta(\omega_n)$, of the eigenvalue condition,  eq.(\ref{spectrum}), is $\eta=-0.25$, for all
 eigenfunctions. The perturbative phase,  $S(0)$, of eq.(\ref{spectrum}), at the infrared boundary $t=0$,  
   is replaced by $S(k_0)$, with $k_0=\Lambda_{QCD} \exp (t_0/2)$ and $k_0 =0.6$ GeV.
 The eigenvalues determined
with and without SUSY effects  differ substantially for $n \ge 3$ whereas for $n < 3$ they show
 no difference. This is understandable from the Appelquist-Carrazone  theorem~\cite{AppCar} because the assumed SUSY threshold, that we have chosen,  lies between
 the critical momenta for the second eigenfunction
  ($ k_c \sim 1$  TeV) and  the third eigenfunction ($k_c \sim 100$ TeV). The $k_c$ values  computed at NLO,
with and without the SUSY threshold, 
are shown in  Fig.~\ref{kcrit}. They turn out to be very close to the leading order values 
  calculated   from eq.(\ref{lokcrit}) - the difference being due to the fact that $\alpha_s$ runs
 more rapidly for NLO than for LO. Furthermore these critical momenta show only  small dependence
  on the presence on  SUSY threshold.


On the other hand,  the eigenvalues, $\omega_n$,
which are important for the description of the HERA structure-function data at low-$x$,
 are very sensitive to possible threshold effects, i.e. they differ substantially already in LO and
  the differences are much larger than any possible uncertainties due to the unknown phase $\eta$.  For example, in LO,  eq.(\ref{omegan-lo}), the ratio of $\betabar$'s below and above the SUSY threshold is 7/3, which means that already for $n \ge 3$  
the effect  of the  change in $\betabar$ on the eigenvalues,  
 is much larger than the maximal possible effects due to the  uncertainty in $\eta$ ($\eta$ can only vary  between $\eta=0.25$ and $\eta=-0.75$).  At NLO, we find that  the phase independent  discrepancy between the eigenvalues with and
 without the SUSY threshold starts at $n \ge 5$ (see Fig,~\ref{omegan}).
    These substantial differences are related to  
 the fact that   for $\omega < 0.1$, the change of  phase of an eigenvalue, $\Delta \phi$,
 arising from the change in $\betabar$ as one crosses the SUSY threshold is large
(as can be seen from  eq.(\ref{deltaphi})) 
 and indeed  much larger than the maximal possible $\eta$ change, $\Delta \eta \le 1$.

\begin{figure}
\centerline{\epsfig{file=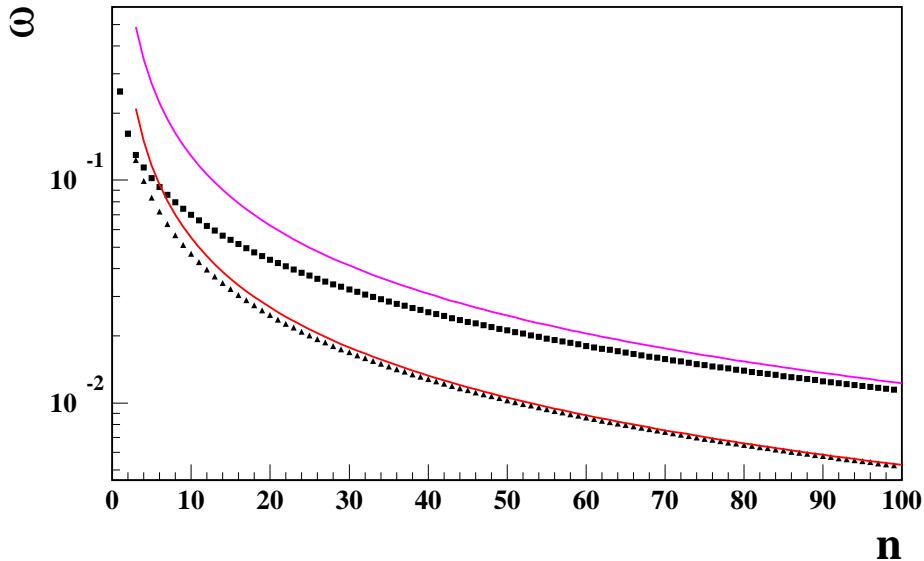, width=14cm}}
\caption{The eigenvalues  computed in the NLO evaluation  of the Standard Model (triangles)
 and SUSY at a threshold of 10 TeV (squares). The lines show the LO computation in the two cases.
   \label{omegall}}
\end{figure}

Since the properties of the eigenvalues are determined by the behaviour at very high
 virtualities,
 (of the order of $k_c$), it should be expected that the eigenvalues computed in NLO
 should approach the LO ones at large $n$ \footnote{ For sufficiently small $\omega$, the NLO effects both in $\betabar$
 and the characteristic function, $\chi$, become negligible.} .
  Fig.~\ref{omegall} shows  the comparison of the eigenvalues computed using the  NLO and LO approximations and  confirms this expectation. The LO computation was made using  eq.(\ref{omegan-lo}) 
with $\betabar$ values computed with $n_f=6$ below  the SUSY threshold of 10 TeV.  It is interesting to observe that LO and NLO results approach each other more slowly  in the case of the SM+SUSY than in the SM alone.
 This  slower approach is due to the fact that $\alpha_s$ runs more slowly  above the SUSY threshold.
This means
 that the eigenvalues, $\omega_n$, approach zero at  a  different pace,
as can be seen from the figure.
We note that for small $\omega$, the
eigenvalues are very closely packed and so the effect of the discrete nature of the solutions becomes less
important. To a good approximation we could replace the sum over the eigenfunctions for
 small
 $\omega$ by an integral
over a  range of small  $\omega$. However, it is important to note that the jacobian for the transition from a discrete
sum to an integral is proportional to the gradient of the $\omega \, - \, n$ distribution shown in
 Fig.~\ref{omegall} and this is  different in the two cases - leading to  different
pomeron amplitudes.

\begin{figure}
\epsfig{file=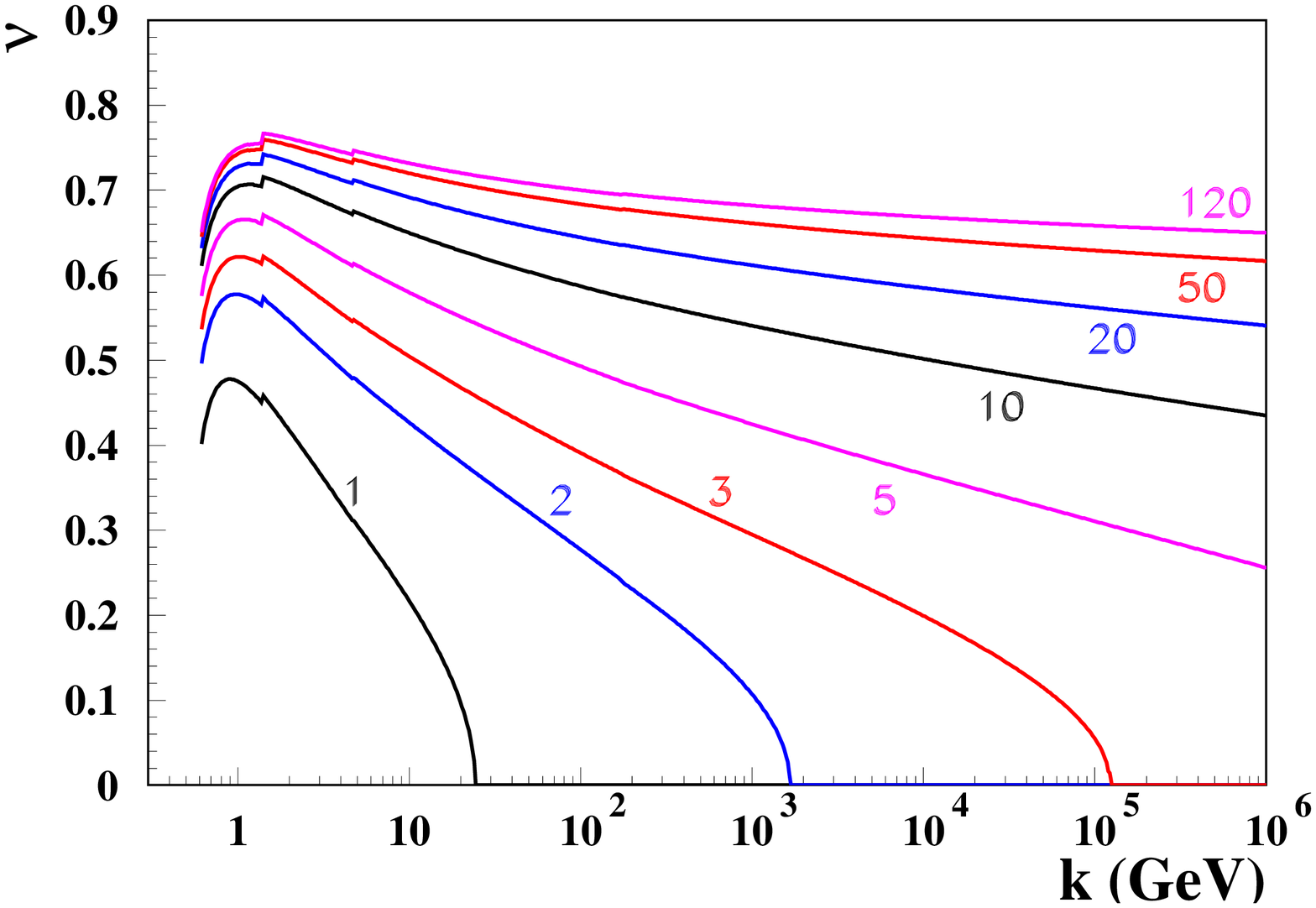, width = 8.5 cm} \
 \epsfig{file=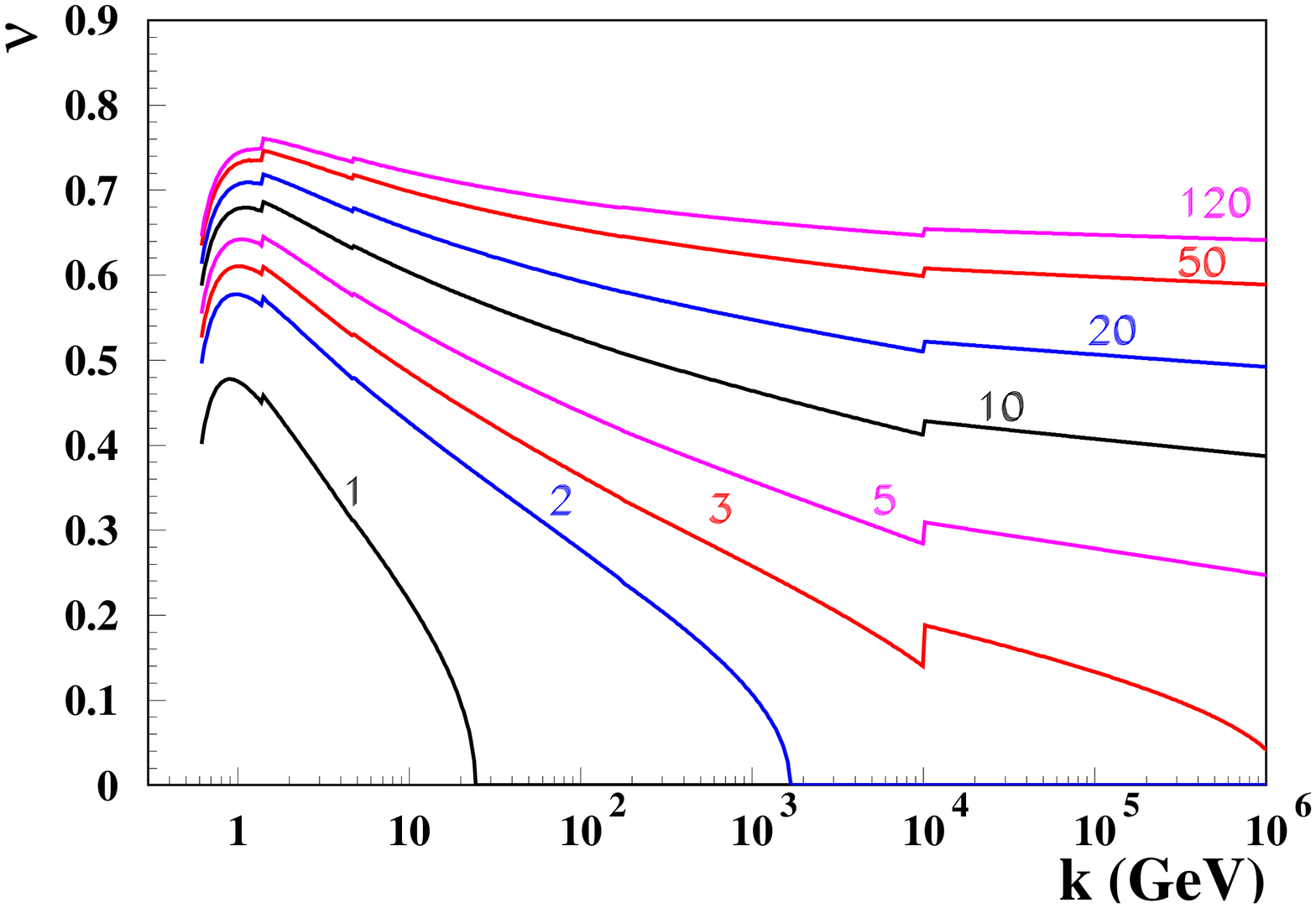,  width = 8.5 cm}
\caption{Oscillation frequencies as a function of gluon transverse
 momentum for various eigenfunctions. The left-hand panel is the case of the Standard
Model and the right-hand panel is the case of N=1 SUSY above a threshold of 10 TeV. For 
the purpose of this comparison it has been assumed that the infrared phases are the same
in both cases.
  \label{nufig}}
\end{figure}

In addition to the change of the running of $\alpha_s$, there  are also  effects due to the  $\delta \chi_1$ contributions to $\chi_1$
which sets in above the SUSY threshold, eq.(\ref{dchif}) and eq.(\ref{dchis}). It is this discontinuity which is responsible 
 for the discontinuities in  the frequencies $\nu$ at threshold, see Fig.~\ref{nufig}, and {\it not}
the change in the rate of running of the coupling, which  remains a continuous 
function~\footnote{ A similar smaller discontinuity can be seen at around 3 GeV. This corresponds
to the c-quark threshold. There are analogous, even smaller, discontinuities
  at the b-quark and t-quarks thresholds}.
The change in frequency thus compensates for the change in the characteristic
function in order to ensure that the eigenvalues, $\omega_n$, remain unchanged as one passes
through the threshold \footnote{The discontinuous changes in frequency are due to the fact that
the change in characteristic function is imposed at a threshold in its entirety. A determination
 of the NLO characteristic function which accounted for the mass of internal particles
would smooth out these discontinuities.}.
\begin{figure}
\centerline{\epsfig{file=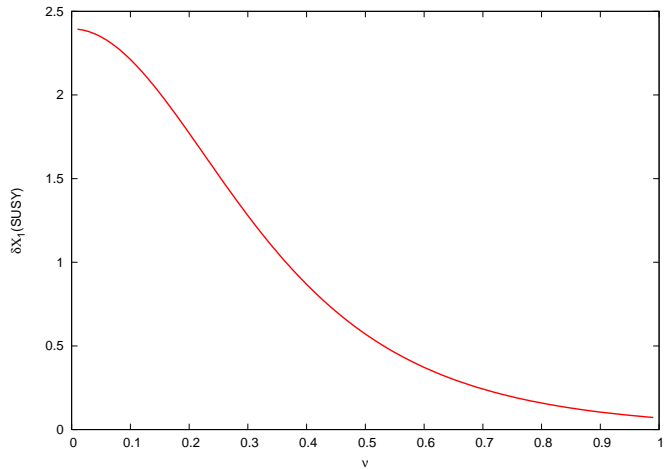, angle=270, width=9cm}}
\caption{ Decrease in the NLO characteristic function, $\chi_1$ as
a function of frequency $\nu$ \label{dchi1}}
\end{figure}
The contribution, $\delta \chi_1$,  of these additional terms 
 is shown as a function of frequency in Fig. \ref{dchi1}
where it can be seen that this is a rapidly decreasing function, which explains why the
discontinuities in frequency 
 at threshold are much larger for the lower eigenfunctions for which the frequency at threshold is 
lower.

\begin{figure}
\centerline{
\epsfig{file=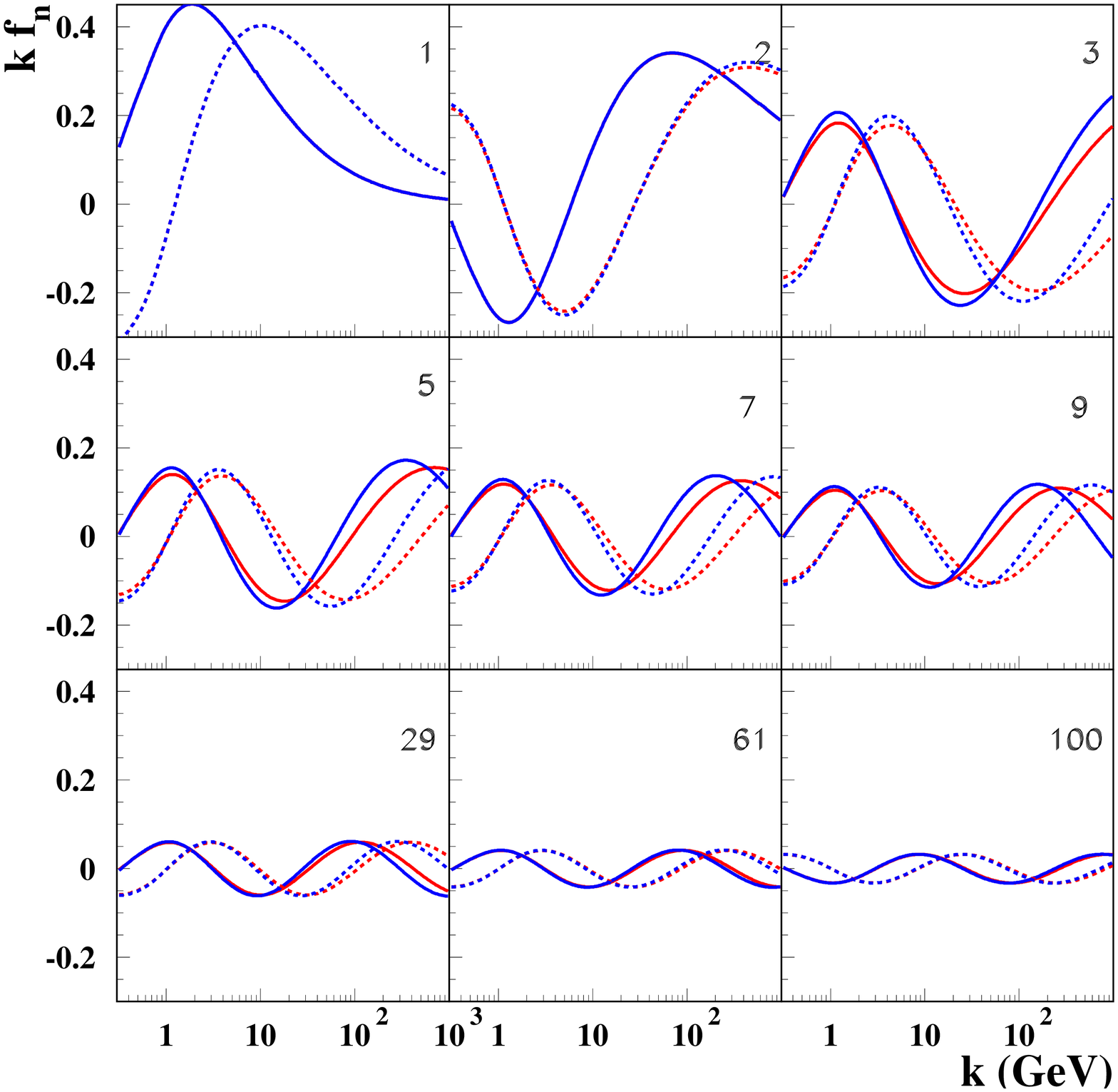, width=14cm}
}
\caption{Comparison of  a representative subset of eigenfunctions
 in the Standard Model (blue) and the SUSY model (red) computed at  $\eta=-0.25$ (solid line) and $\eta=+0.25$ (dashed line). The SUSY threshold is assumed at 10 TeV.  The eigenvalue number is given in the upper right corner.  \label{phiall}}
\end{figure}

For lower $n$ eigenfunctions, the change of the frequencies due to the SUSY threshold leads also to the change of its shape. In Fig. \ref{phiall} we show a representative subset of  eigenfunctions in the Standard Model  and the SUSY model in the transverse momentum region relevant for a fit to HERA data.  
The eigenfunctions are shown with  values of $\eta=-0.25$ and $\eta=+0.25$
  (in order to give an indication of the sensitivity of the allowed
eigenvalues to the unknown infrared phases), 
 with and without  SUSY at a threshold of 10 TeV. 
As expected, the first two
eigenfunctions are identical since their values of $k_c$
are below the SUSY threshold. The third and higher eigenfunctions display significant
differences for both displayed $\eta$ values. 
Remarkably, these differences diminish for higher eigenfunctions 
and for $n>50$ the two eigenfunctions almost overlap in the 
displayed $k_T$ region (relevant for a fit to HERA data). The reason
for this can be seen from Fig. \ref{nufig}, which shows that for the 
relatively low transverse momenta the differences in the frequencies between the
two models decreases with increasing eigenvalue number, so that if the 
infrared phases are  equal, the functions will be almost identical in this region.

 In summary  we can state that the Discrete BFKL Pomeron shows a clear sensitivity to 
      BSM physics effects and that these effects cannot be absorbed into
 its only free parameters, the infrared phase $\eta(\omega)$. This is  clearly seen in the eigenvalue  dependence on the  SUSY thresholds, both in the LO analytical approach, eq.(\ref{omegan-lo}) and eq.(\ref{deltaphi}), and in the NLO numerical evaluation, Fig.~\ref{omegan}. In addition, the asymptotic behaviour of the eigenvalues with increasing $n$  (or decreasing $\omega$)   
  is very different for SM and SM+SUSY  irrespective of the possible higher order QCD corrections, Fig.~\ref{omegall}.  
   This means that we have here  a very different situation 
 from the scenario described by the  decoupling theorem \cite{AppCar} where the large
logarithmic corrections {\it can}  be absorbed into unphysical renormalization constants 
leaving only higher dimension operators whose coefficients are suppressed by powers of
the new-physics mass scale. 
In the case of the DP   the effects of  SUSY thresholds produce large changes of frequencies, Fig.~\ref{nufig}, which
 modify the infrared perturbative  phases $\eta$. These  alter, in turn,  the spectrum $\omega_n$  and 
hence the properties of the gluon density. Since the gluon density is a measurable quantity the non-perturbative  phases, $\eta$, can also be  measured although only
indirectly. 

\section{Comparison with HERA Data}

\subsection{General considerations}
One of the most important results of the HERA experiments is  the measurement  of the gluon density.
This density encompasses the properties of the pomeron in the sense that
 the same gluon density determines the dynamics of the inclusive $\gamma^* p$  (or $F_2$) and  diffractive processes, in particular the exclusive vector meson production. Several investigations performed in the context of the dipole models~\cite{GBW,MSM,BGK, KT,Forshaw2003,KMW,WK,KLMV}   have shown that the effective intercept of the gluon density measured by the rise of $F_2$ with diminishing $x$, called $\lambda$, is  properly translated by  the  optical theorem,  to the effective intercepts seen in the exclusive vector meson production.    

The effective intercept $\lambda$ measured at HERA varies between $\lambda \approx 0.2$ at $Q^2= 10$ GeV$^2$ to about $\lambda =0.35$ at $Q^2=100$ GeV$^2$, see Fig. 9 of~\cite{KLRW}. The $Q^2$ dependence of $\lambda$ in $F_2$ and in its diffractive counterparts can be well reproduced by the  DGLAP evolution in which the values of $\lambda$ are  almost entirely of perturbative origin.  
In the well-known Donnachie-Landshoff (DL) \cite{DL} picture of the Pomeron the variation of $\lambda$ with $Q^2$ is due to the existence of a hard ($\lambda = 0.4$) and a soft, non-perturbative,  ($\lambda=0.08$) Pomeron which ad-mixtures varies with $Q^2$.    

The properties of the gluon density corresponding to the DP  are determined by the Green function constructed from the discrete eigenfunctions of the BFKL  kernel (convoluted with the proton 
 impact factor). In contrast to the DL Pomeron the DP is composed of infinite many  eigenfunctions with eigenvalues varying like $\omega_n \approx 0.5/n$. The eigenvalues $\omega_n$ are almost entirely of perturbative origin because its only non-perturbative ingredients are the infrared phases $\eta_n$,  which have a negligible importance for larger $n$, as was explained in the previous section. 

The infrared phases have, however, a strong influence on the shape of the gluon density
 since they determine how  the contributing eigenfunctions add together.
 Let us  recall that the un-integrated gluon density from the DP is of the form
 \begin{equation}
 \dot{g}(x,k^2)_{DP} \ = \frac{1}{x} \; k^2 \sum_{n=1}^{n_{max}} \left( \frac{k}{x}\right)^{\omega_n} A_n f_n(k,\eta_n)  
 \label{gdens}
 \end{equation}
where $\dot{g}$ means differentiation of the gluon density w.r.t. $\ln(k^2) $.
Here, the eigenfunctions as a function of $k$ (rather than $t$) are normalized w.r.t. $k$ and are related
 to $f_n(t)$ by
 \beq f_n(k,\eta_n) \ = \ \frac{1}{k} f_n\left(\ln\left(\frac{k^2}{\Lambda_{QCD}^2}\right)\right) \label{ttok}. \eeq
$A_n$ is the overlap integral of these eigenfunctions with the proton impact factor, eq.(\ref{protff}),  with the eigenfunctions $f_n(k,\eta_n)$ computed with a specific $\eta - n$ relation, $\eta_n$
(after accounting for the non-zero overlaps of the eigenfunctions
 of the non-Hermitian kernel, -  for details see~\cite{KLRW}).
 The sum over eigenfunctions  in eq.(\ref{gdens}) is limited for numerical reasons (see below) to $n_{max} = O(100)$\footnote{We have shown in ref.\cite{KLRW}  that  an adequate description of HERA $F_2$ data requires O(100) eigenfunctions. Nevertheless, the limit of $n_{max} \sim 100$  represents a model assumption which is sufficient for HERA data but which could be too low for evaluation of LHC DY data. For LHC data it
 could  also  be  necessary to include the contributions of negative $\omega$'s, eq(\ref{nom}).}.  The oscillation frequencies of the eigenfunctions, at transverse momenta relevant to HERA,
vary very little from one eigenfunction to the next, so in order to obtain 
 a positive gluon density, which  grows with $k$,
 it is necessary to generate a strong  $\eta - n$ (or equivalently $\eta - \omega$) dependence;
 the eigenfunctions oscillate in $\ln k$ and the only way
 to cancel these  oscillations is to introduce a shift of the phase between the different
 eigenfunctions.

 The sum of eq.(\ref{gdens}) determines the evolution properties of the gluon density in agreement with the  BFKL equation (\ref{evolut1}). The infrared phases are  determined by the dynamics of  non-perturbative QCD and it should be  possible, in principle, to estimate them  using lattice calculations. It should  also be possible to determine them directly from data at comparatively small $Q^2$. 
  However,
 in order to be able to extract these phases accurately,
 the required data set should have a much larger $x$ and $Q^2$ range than the presently available HERA measurements.  In addition the data set has to achieve the precision of the present HERA $F_2$ data. 

Therefore at present, to be able to confront the DP with data,  we have no other choice but to construct a heuristic model for the infrared boundary condition based on the general understanding of the  non-perturbative physics. We 
first  postulate  that the form of the infrared phase function, $\eta(\omega)$, dictated by the infrared behaviour of QCD, is {\it not} sensitive to  BSM physics and is a smooth function of $\omega$ rather than an arbitrary number for each
eigenfunction. We  then  choose to describe it in terms of a suitable parameterization,  motivated by a similarity of the BFKL dynamics with the Schr\"odinger equation, described in detail below. In addition to the phases,
 we have also to specify the proton 
 impact factor.  

 We consider this heuristic approach as a first step towards the determination of the infrared phases, which  are important quantities of  the non-perturbative QCD.   
 The main purpose of the investigation  reported in ref.~\cite{KLRW}  was to check whether a physically plausible  boundary condition  provides a good description of data, 
i.e whether the Discrete BFKL Pomeron can describe the dynamics  of the measured gluon density. 
 The main purpose of the present investigation is to find out whether
 the genuine sensitivity of the DP to the presence of BSM physics at  high-energy 
 can improve or worsen the quality of the fit to data, notwithstanding the uncertainties
 associated with the infrared phase conditions.

\subsection{The Infrared Boundary}

Our heuristic model of the infrared boundary consists of a set of  physically well motivated assumptions about the proton 
 impact factor and about the $\eta - n$ (or equivalently $\eta - \omega$) relation. 

The proton  
impact factor has to be  positive everywhere and concentrated at the values of $k< {\cal O}(1)$ GeV.   We therefore choose  a very simple possible form 
\beq \Phi_p(k) \ = \ A k^2 e^{-b k^2}, \label{protff}  \eeq 
 as in ref. \cite{KLRW}. We  have also investigated other forms of the proton  impact factor, e.g. with different powers of $k^2$ in the prefactor and/or the exponent but found  that the fit to data has no sensitivity to such alternatives.
 This is  due to the fact that  all eigenfunctions have a similar, oscillatory,  shape near the infrared boundary and that the period of oscillations of the eigenfunctions is much larger than any physically possible support of the proton impact factor.   
Note that the form~(\ref{protff})  vanishes  as $k^2$ for small $k$,
 as required by colour transparency 
 and that the coefficient $b$ has the interpretation of the average inverse
 square transverse momentum of partons inside the proton  (the value of the parameter $b$ was left  though completely free in the fit). 

Our choice of   ansatz for the dependence of the infrared phases, $\eta_n$, 
on the eigenfunction number, $n$,  is motivated from an examination of 
 eq.(\ref{omegan-lo}) for the eigenvalues at L.O. We see that for large $n$ we have
 $\omega_n \, \propto \, 1/n$ - the eigenvalues decrease and become closely packed
as $n$ increases. This is similar to the eigenvalues of a bound state in a Coulomb potential 
problem. The value of $\eta_n$ has a restricted range (in order to avoid ``cross-over''
between adjacent eigenvalues) and its variation with $n$ must be smaller than $\pi$. Since they
are  generated by the quasi-bound states of gluons inside the proton they should be
 described by a simple parameterization. 
In ref.\cite{KLRW}, we found a simple,  two-parameter, form 
\beq \eta_n \ = \ \eta_r \left(
 \frac{(n-1)}{(n_{max}-1)}\right)^\kappa,   
  \label{etan} \eeq
where $n_{max}$ is the number of eigenfunctions we use for the fit and
$\eta_r$ represents the total range (in units of $\pi$). The parameter $\kappa$ must be less than 
one. (Note that for $n_{max} \rightarrow \infty $ and fixed $n$,  the phase  $\eta_n$ formally tends to 0.)

Eq.(\ref{etan}) is by no means unique and we could have added terms which are analytic in
 $\omega_n$ of the form
 $$ b + c \omega_n + d \omega_n^2 + \cdots .$$ 
We have tested such more general parameterizations and found that, despite the introduction
of extra parameters, there is no improvement in the quality of the fit obtained. We therefore
use the simple ansatz (\ref{etan}), but we treat $\eta_r$ as a free parameter (with
the only restriction that it must not exceed one), in order to assure a bias-free evaluation
in all of the fits that we perform.


In ref.\cite{KLRW}, we defined the infrared boundary as a phase condition at the lowest possible  value of the transverse momentum, $k=k_0$, which can be safely reached by the perturbative calculation. To make this value as close as possible  to $\Lambda_{QCD}$ we  considered only the one-loop running of the coupling. This gave a value of $k_0=0.3$ GeV, which corresponds to $\alpha_s \sim 0.7$. The reason for running the coupling at one-loop only  was that in principle this is the same order of perturbation
theory as the NLO characteristic function, $\chi_1$ \cite{FL}.
However, given that we modify the eigenvalue eq.(\ref{hr}) by resumming all the large
corrections in $\chi_1$ using the technique of ref.\cite{salam}, 
it is more appropriate to take the $\beta$-function  to two-loop order
which is what we use in this paper.

When we do this, we are faced with a problem - namely that we cannot run the coupling
below an ``infrared'' scale $k_0=0.6$ GeV, which corresponds to $\alpha_s\sim 0.7$ 
(at the two loop level), without approaching the Landau pole too closely.
On the other hand, the infrared boundary conditions are to be imposed at a transverse momentum of order $\Lambda_{QCD}$.
 Moreover, we need to know  the eigenfunctions below $k_0$ in order to
perform a convolution with the proton impact factor, which has support mainly below
 $k_0$. Therefore, guided by the behaviour of the eigenfunctions in 
the  perturbative region,
 we continue them  down to a lower momentum $\tilde{k}_0$, which should be of order $\Lambda_{QCD}$,
  using the extrapolation
of the phase  $\phi_n(k)$ 
\beq \phi_n(\tilde{k}_0) \ = \ \phi_n(k_0) - 2 \nu_n^0 
\ln\left(\frac{k_0}{\tilde{k}_0}\right),
\label{eq:phrel}
 \eeq
where for each eigenfunction, with index  $n$, $\nu_n^0$ is the frequency of the oscillations
near $k=k_0$~\cite{KLRW}. We have assumed that this frequency is constant below $k_0$, an assumption
which is correct for sufficiently small $k_0$,
at least for the leading order BFKL kernel (see \cite{lipatov86}). 
Any deviation from constant frequency should have a negligible effect as we are only extrapolating over a small range in gluon transverse momentum. The numerical values of $\nu_n^0$  are obtained by inverting the eigenvalue equation (\ref{hr}), modified according to \cite{salam}. 

The overlap integrals between the proton impact factor and the eigenfunctions must also start
at $\tilde{k}_0$ (the support of these impact factors being significantly attenuated
at $k_0$). We therefore use this momentum at which we impose the infrared phases of 
the eigenfunctions. The relation between the phases at $k_0$ and $\tilde{k}_0$ is given by
 eq.(\ref{eq:phrel}). We leave the exact value of $\tilde{k}_0$ as a free parameter with the
restriction that it must be ${\cal O}(\Lambda_{QCD})$ and define it to be the scale
at which the phase of the leading eigenfunction vanishes (as can be seen from eq.(\ref{etan})).

\subsection{ Results of the fit}

Before a comparison can be made with the measured structure function, $F_2$,
 it is necessary to convolute the gluon density with the impact factor for the virtual photon
 (for  details of the procedure  see Section 6 of ref.\cite{KLRW}). The impact factor for the virtual photon is
calculable in perturbative QCD and has support which is peaked at transverse
momenta of the order of the photon virtuality, $\sqrt{Q^2}$.


The fits were performed using the HERA data \cite{H1ZEUS} in the low-$x$ region, $x \, < \, 0.01$.
To avoid any saturation effects we have limited the fit to the  $Q^2 \, > 8 \  \mathrm{GeV}^2$ region. We recall that the saturation scale at HERA was determined to be $Q^2_S = 0.5 $ GeV$^2$ at $x\approx 10^{-3}$~\cite{KMW,WK}, therefore our choice of the $Q^2$ region is very conservative.   
This choice  means that, in this paper, we concentrate on the one-pomeron exchange,
 without any multi-pomeron contributions, which could  induce saturation effects. The saturation effects could also play a role without multi-pomeron effects through a modification of the boundary conditions, see ref.~\cite{Muel-Triant}, which in turn could modify our ansatz for the infrared boundary. In any case our choice of the   
 $Q^2$ region for fits  assures that saturation effects can be ignored in this first evaluation. In  future  we plan to extend our analysis into the $Q^2$ region which could be more sensitive to saturation.   
   
 In the region of  $Q^2 > 8$ GeV$^2$ we have a total of 108 data points and a total
of 5 parameters - so the number of degrees of freedom is $N_{df}= 103$. We consider the $Q^2 \, > 8 \  \mathrm{GeV}^2$  region as our main investigation region and use the 
$Q^2 \, > 4 \  \mathrm{GeV}^2$ as a cross check.

As discussed in the previous paper \cite{KLRW}, in order to obtain the most accurate 
estimate of the un-integrated gluon density
 we should include in the fit  as  many of the higher $n$ eigenfunctions as possible. 
 Indeed, we observe that the fit quality improves  with increasing number of included eigenfunctions and the series  converges in $\chi^2$. In principle this convergence should improve as $n \rightarrow \infty $.
 However, in practice  the number of eigenfunctions used in a fit is  limited by the
  numerical precision of our calculation. 
We have indications  that if we take significantly more than 100 eigenfunctions our fit could be polluted by numerical instabilities arising from an accumulation of computational rounding errors. 
Moreover, we find no  improvement
in the quality of our fits, either in the case of the Standard Model or for MSSM SUSY at any
of the thresholds investigated, when the maximum number of eigenfunctions, $n_{max}$,
exceeds 100. We have therefore taken $n_{max}=100$ throughout.

\begin{table}
\begin{center}
\begin{tabular}{||c|c|c|c|c|c|c||}  \hline 
\begin{tabular}{c} SUSY Scale \\ (TeV) \end{tabular}
 & $\chi^2$ & $\kappa$ & $\tilde{k}_0 \ (GeV)$ & $\eta_r$ & A & b (GeV$^{-2}$) \\ \hline \hline
 3 & 125.7 & 0.555 & 0.288 & -0.87 & 201.2 & 10.6  \\  \hline
 6 &      114.1 &    0.575  &   0.279 &   -0.880 &     464.8 &      15.0 \\ \hline
 10 &   109.9 &   0.565 &  0.275 & -0.860 &   693.1 &   17.4  \\ \hline
 15 &   110.1 &  0.555 &  0.279 & -0.860 &   882.2 &  18.6 \\ \hline
 30 &  117.8 & 0.582 &  0.278 & -0.870 &  561.6 &   16.2 \\ \hline
 50 &  114.9 & 0.580 &  0.279 &  -0.870&  627.4 &   16.8 \\ \hline
 90 & 114.8  &  0.580  & 0.279 &  -0.870  &  700.2 &   17.5 \\ \hline
 $\infty$   &      122.5  &   0.600 &  0.294 &  -0.795  &   813.1  &   18.2 \\
\hline \hline
\end{tabular}
\caption{Fits for N=1 SUSY at different scales. The bottom row corresponds to
the Standard Model. All fits are performed with $n_{max}=100$. } \label{table8}
\end{center}
\end{table}

In Table \ref{table8} we show our fits for various SUSY thresholds as well
as the Standard Model. Let us first note that  the  $\tilde{k}_0$ values obtained in the unbiased fit, $\tilde{k}_0\sim 275$ MeV, are close to  $\Lambda_{QCD}$. 
At the same time the  value of $b$ 
 implies that the proton impact factor peaks around $\Lambda_{QCD}$, as 
expected for a self consistent description. This together with the relatively low $\chi^2$'s of all fits confirms the success of  our construction of the infrared boundary.



The quality of the fits shows a clear preference of the evaluation with SUSY effects; the fit for the Standard model
is  worse than the fits with SUSY thresholds larger than 3 TeV.
A SUSY threshold of 3 TeV, which is close to the reach of
LHC also gives  a worse fit. On the other hand for a SUSY threshold in the
region of 10 - 15 TeV, the quality of the fit is 
significantly improved, but that for significantly larger SUSY thresholds
 worsens again. \\

Let us note that the differences between the $\chi^2$ fits shown in  Table \ref{table8} are very significant because the Maximum-Likelihood method, which assures that the minimum of  $\chi^2$ provides the best estimate of the parameter values, also states  that 
$1\sigma$ error in determination of  parameter values is given by $\Delta \chi^2 =1$,  irrespective of the number of degrees of freedom. Therefore, the differences between the fits of Table~\ref{table8} are a multi-$\sigma$ effect, within our model of the infrared boundary. 

The $\Delta \chi^2 =1$ rule is valid as a an estimate of the parameter error only for estimates within one theoretical framework, i.e. for one likelihood function. On the other hand,
when $\chi^2$ is used to quantify the agreement of different theories with data, it is expected that the observed $\chi^2$  can deviate from the optimal value because it fluctuates with the  probability density function, $f(\chi^2,N_{df})$, which  is approximately a Gaussian with  the average value equal to $N_{df}$ and the variance, $\sigma^2=2N_{df}$. In the case of fits, presented in the Table 1, the expected  $\chi^2$ should be around 103
 and  $\sigma=14$. Therefore the $\chi^2$ values obtained for best fits with the
 SUSY mass ${\cal O}(10)$ GeV lie well within one standard deviation. The DGLAP fits  have a $\chi^2/N_{df} \approx 0.95$. This would lead for our sample to $\chi^2=98$, 
which is also within one standard deviation of the optimal value, 
 so that one cannot conclude that either fit is better.
 The evaluation with the {\it goodness of fit} criterion   called {\it p-value},
 which is more appropriate for $N_{df} \sim 100$, gives a {\it p-value} $\approx 30\%$
 for SUSY masses of ${\cal O}(10) $ GeV, which is again an excellent result,
 see ref.~\cite{Caldwell,Blobel}. \\

As a check, we  also performed the fits with a  lower $Q^2$ cut,  $Q^2 \, > 4 \  \mathrm{GeV}^2$. 
We find that in this $Q^2$ region there is an significant increase of $\chi^2/N_{df}$ presumably due to various higher order effects, such as the NLO contribution to the photon impact factor or valence quarks effects and possibly also the proximity
of the saturation region.
Although the overall quality of the fit for all data with $Q^2 \, > \, 4   \, \mathrm{GeV}^2$
is significantly worse than with $Q^2 \, > \, 8  \, \mathrm{GeV}^2$
the preference for  N=1 SUSY with the threshold  region of 10-15 TeV 
is also clearly seen. 
In the $Q^2 \, > \, 4  \, \mathrm{GeV}^2$ region there are 128 points and the $\chi^2$ 's of the best fits are 184.3 (3TeV),  164.5 (6TeV),   155.6 (10TeV), 152.6 (15TeV), 169.7 (30TeV), 164.7 (50TeV), 164.3 (90TeV). The best $\chi^2$ for Standard Model is 169.7.  The values of the fit parameters are similar to the values shown in Table \ref{table8}, for the $Q^2 \, > 8 \  \mathrm{GeV}^2$ region.

\begin{figure}
\centerline{\epsfig{file=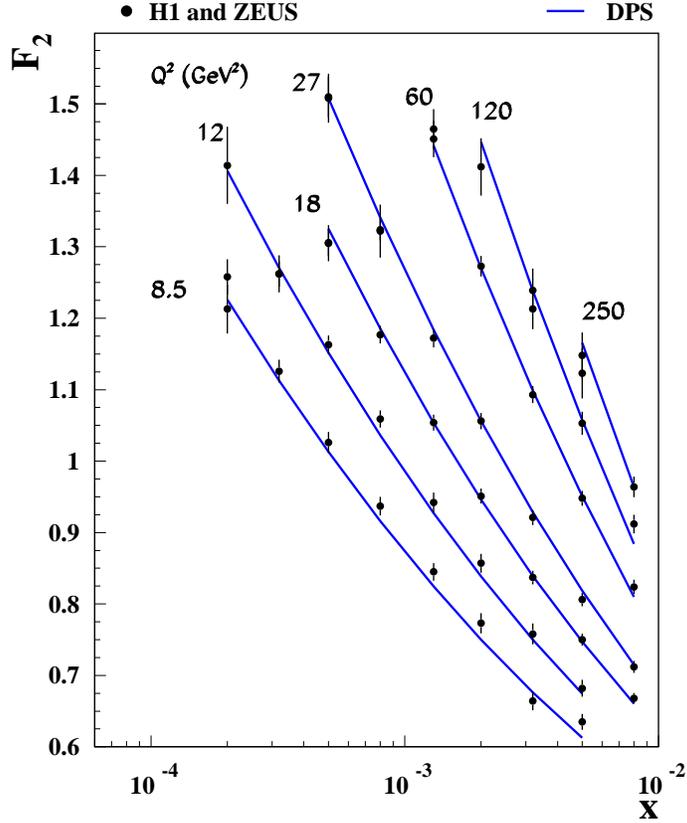, angle=0, width = 10 cm}}
\caption{ Comparison of the DP fit with $M_{SUSY}=10$ TeV with HERA data. } 
\label{globfit}
\end{figure}

\section{Discussion}

The main result of this paper - namely a possible sensitivity of HERA data to BSM effects is especially astonishing as it was not seen in the usual solutions to the BFKL equation~\cite{AMKS,KMS,CCS}. 
 This sensitivity  is due to the fact that we use the Green function method,
in which the Green function is considered to be a universal property of the BFKL equation
and that apart from supplementing this equation with a given set of infrared boundary conditions
for the eigenfunctions, no cuts on transverse momentum are applied, so that
    the conformal invariance is broken in a very smooth way, solely  by the running of the coupling constant.

Had we imposed kinematic limits on the gluon virtuality for the individual eigenfunctions, then
for those eigenfunctions with $\omega  \lessapprox 0.1$, they would never reach the region
in which they decay as required for compatibility with a DGLAP analysis in the DLL limit.
Alternatively, one might attempt to fit HERA data only with the first eigenfuction,
which does indeed start to decay within the HERA region, making the assumption that the
higher eigenfunctions (with lower $\omega$) only couply weakly to the proton.
However, the  properties of this first eigenfunction are in  clear contradiction with data;  for example, the rate of rise of $F_2$ with diminishing $x$ would be independent of $Q^2$ with a value of $\lambda \approx 0.25$ whereas the $\lambda$ value in data varies between 0.15 to 0.35 in the observed $Q^2$ region.   The results of ref.~\cite{EKR}  indicate that one needs at least four eigenfunctions to describe HERA data (making an artificial assumption that the overlap constants $A_n$ of eq.(\ref{gdens}) are totally unconstrained), which  extends the virtuality region to $k^{(4)}_c \sim 1000$ TeV.

\begin{figure}
\centerline{\epsfig{file=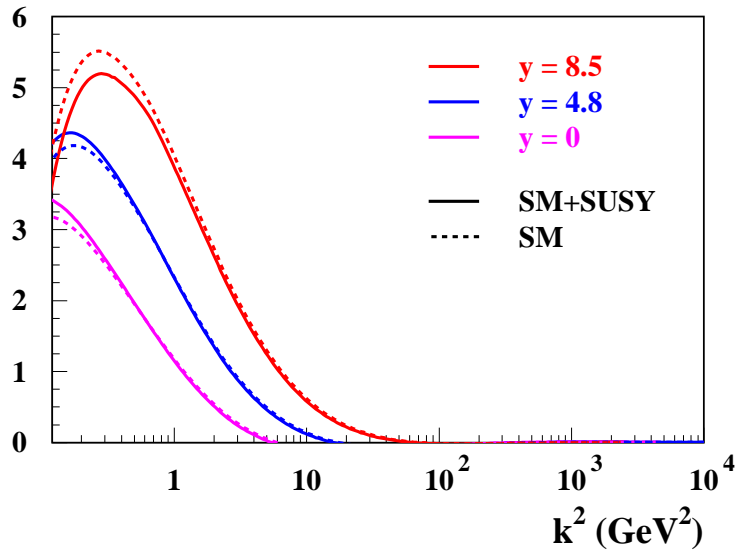,  width=12cm}}
\caption{ Evolution of the initial wave-packet in the DP solution of the BFKL equation as determined from the fit to HERA $F_2$ data. \label{wavepacket}}
\end{figure}


In the solution which we have adopted the momentum conservation is incorporated by convoluting the  Green function with the proton form factor as defined by the solution of the BFKL eq. (\ref{evolut1})
\beq {\cal WP} (y,k^2) \ = \
\int d\omega \int \frac{dk^{\prime \, 2}}{k^{\prime \, 2}}  
\, e^{\omega y} \,
 \hat{{\cal G}}_\omega(k,k^\prime) {\Phi}_p(k^\prime),
 \label{wpevol} \eeq
where in analogy to eq.(\ref{gdens})  the Green function, $\hat{{\cal G}}_\omega(k,k^\prime)$,
 is written in terms of the eigenfunctions $f_n(k,\eta_n)$ and $f_n(k^\prime,\eta_n)$. This expression describes the evolution of a  wave-packet, ${\cal WP}$, from  rapidity $y=0$  to the larger rapidity $y$ values~\footnote{note that the analogy to the QM wave-packet is not complete because in the BFKL eq. rapidity is analogous  to
 imaginary  time.}. 
The (approximate) momentum conservation emerges here  because 
 the quasi-local nature of the kernel ${\cal K}(t,t^\prime)$ ensures that there is 
   no evolution into the very large transverse momenta, apart from the usual BFKL diffusion, $\ln k_T \sim \sqrt{\alpha \ln s}$.  Fig.~\ref{wavepacket}  shows 
  the wave-packet   as it evolves   from  the smallest to the  largest  rapidity  values of the  HERA region,  $y=\ln(1/x)$, and for the SM and SM+SUSY cases.
  
  The initial wave-packet, the curve ${\cal WP}(y=0,k^2)$ shown  in Fig.~\ref{wavepacket},   is not quite the same as the function $\Phi_p(k)$ defined in 
eq.(\ref{protff}) (divided by $k^2$)
 but rather it is related to it by eq. (\ref{wpevol}), i.e. 
it is  built only out of  the eigenfunctions which obey the imposed boundary
conditions on the Green function.
For large $y$, it is sufficient to consider only the discrete eigenvalue part of the Green function
 in eq.(\ref{wpevol}), and
so we are actually using that part of the proton impact factor which is orthogonal to
the (continuum) negative $\omega$ eigenfunctions. 
The initial wave-packet 
is somewhat broader than the  distribution of eq.(\ref{protff}) but it has
the required feature that it is localized to a  transverse momenta of $ \sim $~1~GeV. 
With increasing rapidity it  broadens   due to the BFKL diffusion but this broadening remains moderate, of a size of just few GeV,    as in the usual solutions of the BFKL equation.
   Although our  kernel is constructed from the oscillating eigenfunctions with properties determined by the BFKL dynamics at very high transverse momenta, the oscillations cancel away due to the choice of the phases $\eta_n$ in the fit procedure  and what remains is only a slight broadening of the gluon diffusion spectrum with increasing $y$.


\begin{figure}
\centerline{\epsfig{file=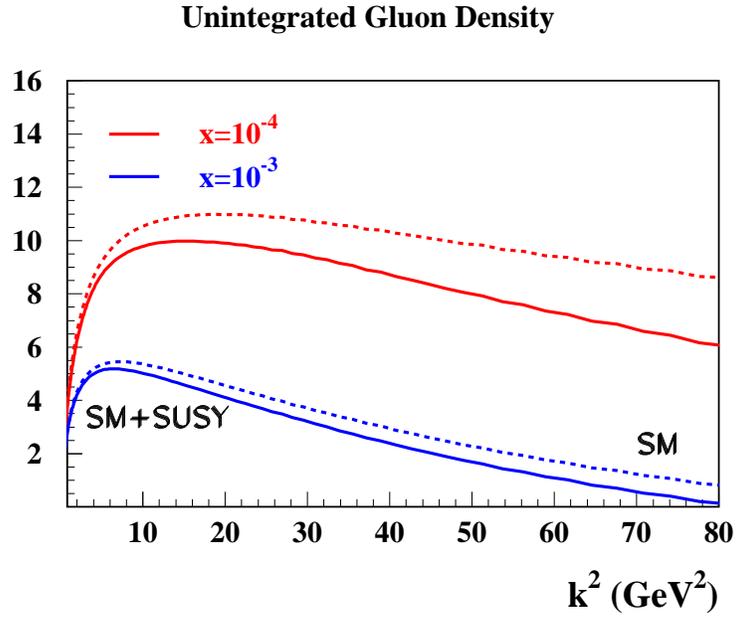,  width=12cm}}
\caption{ The unintegrated gluon density using the DP solution of the BFKL equation,
 as determined from the fit to HERA $F_2$ data. \label{gluden}}
\end{figure}

\begin{figure}
\centerline{\epsfig{file=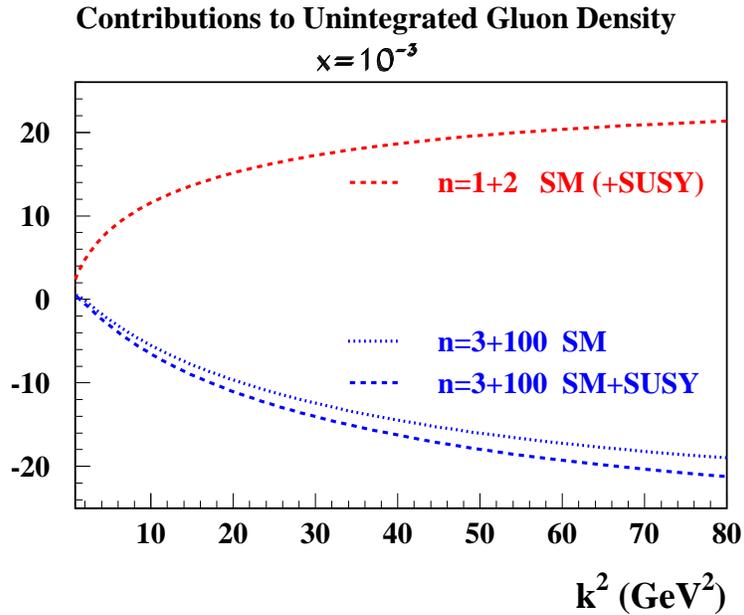,  width=12cm}}
\caption{Comparison of contributions to the unintegrated gluon density from
 the eigenfunctions which are not sensitive ($n=1,2$) and  which are sensitive to BSM effects ($n=3..100$) to the curve at
 $x=2 \times 10^{-3}$ of Fig.~\ref{globfit} \label{greenfcon}}
\end{figure}

In the application to DIS it is more customary  to consider the evolution of the unintegrated gluon density 
  given by 
 \beq x \dot{g}(x,k^2) \ = \  k^2
\int d\omega \int \frac{dk^{\prime \, 2}}{k^{\prime \, 2}}  
\left(\frac{ k x}{k^\prime}\right)^{-\omega}
 \hat{{\cal G}}_\omega(k,k^\prime) {\Phi}_p(k^\prime),
 \label{ugug} \eeq
 which differs  from the evolution of the wave-packet, eq.(~\ref{wpevol}), by the kinematical factor $k^2$ and the factor $(k/k')^{-\omega}$. The latter factor encodes the difference beween Bjorken-$x$ 
and rapidity, $y$. It
  is close to one and  of minor importance. On the other hand,
 the  factor of $k^2$, is important as it amplifies the high virtuality  contributions to the structure function (note that $F_2$ is obtained from a convolution of the unintegrated gluon density with the photon impact factor which selects $k^2$ values close to the $Q^2$ of the experiment).    
 
In  Fig.~\ref{gluden} we show 
the unintegrated gluon density  at two values of $x$ -  one in the middle of HERA region,  
$x=10^{-3}$, and one just above HERA  low x region,  
$x=10^{-4}$. The Figure shows again that the contribution of BSM physics changes  
the shape  of the gluon density but does not lead to an increase of the  
large transverse momentum  tail (in fact this tail turns out to be  smaller for SM+SUSY)  
contrary to the naive expectation that BSM effects should show up as  
an increased parton activity at large virtualities. 
The figure  also shows that the discrepancy between the unintegrated gluon density for the SM
alone  and for SM+SUSY  increases substantially with diminishing $x$ at larger $k^2$, which suggests 
that Drell-Yan measurements at  LHC, in the region of $x \, \sim  \,  10^{-4}$ and $k^2 \sim 50$ GeV$^2$, could  
have high sensitivity to BSM physics.

  In contrast to the   diffusion effect, the important point that we are stressing is that it is
the spectrum of allowed eigenvalues (and also the shape of the eigenfunctions) that
enters into the construction of the Green function  which is sensitive to physics at high scales. 
This sensitivity emerges from the  interplay of two main features of the DP solution: the identity of the boundary condition in respect to  addition of $n \pi$ to the phase and the running of the coupling constant. 
For  $n\ge 3$ this interplay is affected by  SUSY effects,
 whereas for $n\leq 2$ the eigenvalues are unaffected since the critical transverse
 momentum, $t_c$ is below the SUSY threshold.
 For $n \ge 3$, the change of the eigenvalues is substantial.
 As $n$ increases, the ratio of eigenvalues with and without SUSY  rapidly 
reaches its asymptotic value of 7/3. 
In Fig.~\ref{greenfcon} we compare the summed
 contribution of the eigenfunctions which are sensitive to BSM effects ($n=3...100$) with the sum of  contribution of the first two eigenfunctions which are not sensitive to SUSY.
 The contributions of the $n=3...100$ eigenfunctions changes both in size and shape 
when evaluated with and without SUSY whereas the contribution of the first two eigenfunction remains
 unaltered. 

\begin{figure}
\centerline{\epsfig{file=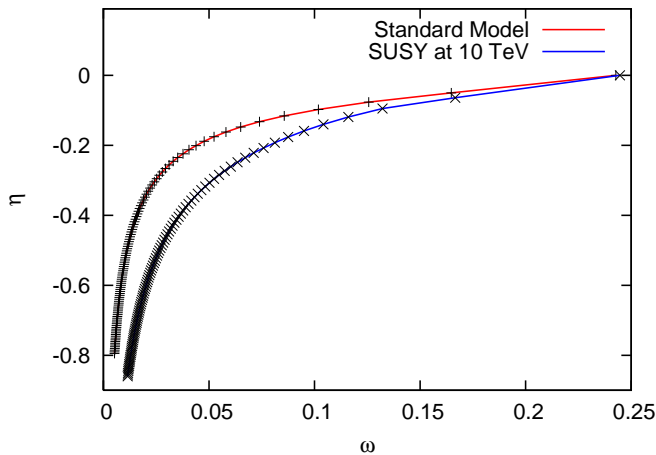, angle=270, width = 9 cm}}
\caption{ The eigenvalues and infrared phases for the Standard Model and  N=1 SUSY at 10 TeV, as determined at $k=\tilde{k}_0$. \label{etaomega} }
\end{figure}

   Fig.~\ref{etaomega} shows the $\eta$ variation as a function of the eigenvalue, $\omega$, for the fit with the SUSY threshold of 10 TeV and for the SM fit only. 
Both relations show a substantial variation of the phase $\eta$ with decreasing $\omega$.  The substantial difference between the two relations reflects a large difference between the eigenvalues and eigenfunctions in  both cases.

The changes due  to BSM effects 
  lead to a substantial increase of $\chi^2$ when the corresponding gluon densities
 are confronted with  data because the changes are of the order of 10\%
  whereas the data precision is around 1 - 2\%. Indeed,    
  the  evolution of the unintegrated gluon density
 performed with the SM alone - using the same  parameter values for $\eta$ boundary  as in the SUSY case
  (dotted line in Fig.~\ref{greenfcon},
 dashed lines in Fig.~\ref{gluden}) would give an increase   of $\chi^2$ by  $ \Delta \chi^2 \sim 160$ when compared to data. On the other hand, had we used the $\eta$ boundary parameters as determined with the SM alone and performed the evolution with the SM+SUSY  eigenvalues and eigenfunction,  the increase of $\chi^2$ would be even larger, $\Delta \chi^2 \sim 300$.  The differences between the $\chi^2$'s of the  SM and SM+SUSY fits shown in Table~\ref{table8} are substantially smaller, $ \Delta \chi^2 \sim 12$,   because  the parameters of the $\eta$ boundary conditions  and the proton form factors were fitted to data what  diminishes substantially the sensitivity to BSM effects.

   It is quite  clear that the DP could provide an exciting framework to study BSM effects if more can be learned about its infrared boundary. 
   One way to improve our knowledge of this boundary is to  apply our analysis  to an another,
 independent set of data, such as the Drell-Yan processes at the LHC \footnote{Natural candidates would be also the diffractive HERA processes. Unfortunately, they are not measured with sufficient accuracy because the HERA detectors were not designed to measure diffractive processes.}. The LHC Drell-Yan data extend in the low-$x$ region to much larger $Q^2$ scales than HERA data. This   will allow the study of  the evolution in $\ln k^2$ which is dominated in the DP scheme by the low $\omega$ region. This region and the region of very low $x$ is very sensitive to  SUSY effects
 and at the same time is much less sensitive to NLO effects which are difficult to compute.

 Future HEP experiments  which are now under discussion can also substantially improve the knowledge of the infrared boundary and of the properties of the DP. The LHeC project~\cite{LHeC}  could provide important information about the region of very low $x$ and not so high scales. This could lead to a better understanding of
 the properties of the large $\omega$ contributions. In addition,  it will be also possible to measure precisely the exclusive diffractive processes. This will provide an independent evaluation of the infrared boundary since in the exclusive processes (e.g in the exclusive $J/\psi$ or $\Upsilon$ production) the gluon density contributes almost quadratically~\cite{KT,KMW}.

The properties of the DP can also be studied  very well 
  in the $\gamma^* \gamma^*$ process in a  future linear $e^+e^-$ collider~\cite{LinCol}. The  $\gamma^* \gamma^*$ process is very interesting because it would permit the direct  test of the universality of the BFKL pomeron and its boundary conditions. Of particular interest could also be the newly 
proposed  electron wakefield accelerator~\cite{Caldwell2},  which could  accelerate electrons into the several TeV  energy region. This  would allow the measurement of the $\gamma^* \gamma^*$ structure function  at very low  $x$ and high $Q^2$'s, which could further increase the sensitivity to BSM effects.  

\section{Summary}

 In this paper we have analyzed the properties of the  Discrete BFKL Pomeron (DP) using analytical and numerical methods. We show, using as example N=1 SUSY, that  BSM physics  substantially alters
 its eigenvalue spectrum and the shape of its eigenfunctions.   This is a genuine sensitivity insofar as  it cannot be 
entirely absorbed into any free parameters of the Discrete Pomeron solution of the BFKL equation. 

The physical origin of this sensitivity can be traced back to the fact that in the 
 low Bjorken-$x$ regime
the behaviour of the structure functions is directly related to the positions of the Regge poles (pomerons). In
the BFKL approach, due to the approximate conformal invariance,  the corresponding eigenfunctions and eigenvalues of the BFKL kernel
are determined from exponentially large transverse moments, where contributions from particles
of any BSM physics play an essential role. On the other hand
the locality property of the BFKL equation assures that in any physical process performed at energy scales which are much lower than the BSM ones, the BSM quanta cannot be produced and the transverse momenta of the virtual particles are very limited. The effects of large transverse momenta  appears,
 however, through the  substantial alteration of the eigenstate spectrum of the BFKL Hamiltonian, which is important in the low-$x$ region. In our view, this provides a new mechanism for the detection of BSM effects, which  has not
 previously been considered.

The eigenvalue spectrum of the DP cannot be directly measured because it determines the gluon density through a complicated superposition of Pomeron states. The result of their interference can be compared with data only after the free parameters of the BFKL solution, determining the infrared boundary, are specified. Although the infrared boundary is a physical quantity of  non-perturbative QCD origin,   we could only determine it in this and the previous paper  within a heuristic approach. Our description of this boundary provides a very good fit to the   data and shows that, the BSM effects are sensed by the HERA $F_2$ data,  notwithstanding the large freedom of the parameter choice.   

The analysis of HERA data 
indicates an improved quality of fit for the case of N=1 SUSY, with
 the SUSY scale as being around 10 TeV. Needless to say that this determination is only possible within our  heuristic model approach. Our  limited knowledge of the infrared boundary  diminishes substantially (but not completely) the  sensitivity of the fit  to BSM effects. 

This sensitivity  can be substantially  improved by a better  determination of the universal boundary condition. We can gain a better understanding of the infrared boundary from the analysis  of additional data sets, especially of the LHC Drell-Yan data. 
The data from future experimental facilities like LHeC, the $e^+e^-$ linear collider or even higher energy plasma wakefield accelerators could also become crucial.
The  $\gamma^* \gamma^*$ process which can be very well measured at  linear colliders is of particular interest since in this reaction the properties of the discrete Pomeron solution are simplified owing to the absence of  the proton. 


The method described in this paper opens a new possibility of using high precision experiments
 to search for new physics  at energy scales considerably larger than the scales at which the experiments
  are performed. 
  We consider the  approach formulated here,
which involves a heuristic model for the parameterization of the infrared phases of the
BFKL eigenfunctions,
  as a first step which should be substantially improved by involving more data and more theoretical analysis. 
    
\bigskip

{\noindent}{\bf Acknowledgements:} \\
The authors are grateful to the Marie Curie Foundation for an IRSES grant,  LOWXGLUE Project 22498,
which has facilitated this collaboration. 
We wish to thank
the St. Petersburg Nuclear Physics Institute, and Southampton University, for 
their hospitality while this work
was carried out.

\noindent
We are grateful for illuminating and useful discussions  
 with J. Bartels about the physical interpretation of the discrete BFKL pomeron.
 We would like to thank Al Mueller for a  careful reading of the manuscript and encouragement.   We are also 
grateful
 to  A. Caldwell, A. Geiser, E. Lohrmann and R. Mankel
 for  useful discussions  about the meaning of $\chi^2$ tests.
 We also thank J. Ellis and A. Weiler for lively discussions.

\end{document}